\documentclass[titlepage,12pt]{article}
\usepackage{amssymb,psfig,epsfig}
\begin{document}
\def\shat{\hat{s}}
\def\rshat{\sqrt{\hat{s}}}
\def\sp{{\mathbf s}_1}
\newcommand{\Bsla}{B\!\!\!/}
\newcommand{\Lsla}{L\!\!\!/}
\newcommand{\sm}{{\mathbf s_2}}
\newcommand{\kh}{{\hat{\mathbf k}}}
\newcommand{\ph}{\hat{\mathbf p}}
\newcommand{\dhh}{\hat{\mathbf d}}

\begin{flushright}
PITHA 02/08
\end{flushright}
\vspace{0.8cm}
\begin{center}
{\LARGE {{\bf CP VIOLATION and BARYOGENESIS}
\footnote{Lectures given at
the International School on {\it CP Violation and Related
  Processes,} Prerow, Germany, October 1 - 8, 2000,  
and at the workshop of
the {\it Graduiertenkolleg Elementarteilchenphysik} 
of Humboldt Universit\"at, Berlin, April 2 - 5, 2001.}
 }} 
\par\vspace{2.0cm}
{\large \bf  Werner  Bernreuther}
\par\vspace{1cm}
 Institut f. Theoretische Physik, RWTH Aachen, 52056 Aachen, Germany
\par\vspace{1cm}
{\bf Abstract:}\\
\parbox[t]{\textwidth}
{In these lecture notes an introduction is given to some ideas and attempts
 to 
understand the origin of the matter-antimatter asymmetry of the
universe. 
After the discussion of some
basic issues of cosmology and particle theory
the scenarios of electroweak baryogenesis,  GUT baryogenesis, 
and leptogenesis are  outlined.}
\end{center}
\vspace{2cm}


\section{Introduction}

CP  violation has been observed so far in the neutral K meson system,
both in $|\Delta S|=2$ and  $|\Delta S|=1$ processes, and recently
also in neutral B meson decays. These phenomena  are very
probably caused by
 the Kobayashi-Maskawa (KM) mechanism, that is to say by
a non-zero phase  $\delta_{KM}$ in the coupling matrix of the
charged weak quark currents to W bosons. CP violation  found so far in
these meson systems does not catch the eye: either 
the  value
of the  CP observable
or/and  the branching ratio of the associated mesonic decay mode is  small. 
However, the interactions that give
rise to these subtle effects may have also been jointly responsible for an 
enormous phenomenon, namely for 
the apparent matter-antimatter asymmetry of
the universe. In this context
it has been a long-standing question
whether or not   CP violation in $K^0
- {\bar K^0}$ mixing, i.e. the parameter $\epsilon_K$, 
is related to the baryon asymmetry of the
universe (BAU) $\eta = (n_b -n_{\bar b})/n_{\gamma} \sim 10^{-10}$. 
In particular, is the experimental result ${\rm Re}\  \epsilon_K >0$
related to  the fact that our universe is filled with matter rather
than antimatter? Because the CP effects  observed so far
in $K$ and $B$ meson
decays are consistently explained by the KM mechanism, one may 
paraphrase these questions in more specific terms by asking whether 
 the standard model of
particle physics (SM) combined  with the
 standard model of cosmology (SCM) can explain
the value of $\eta$? This  has been answered in recent years
and, surprisingly, the answer does not refer  to the
role the KM phase $\delta_{KM}$ may play in these explanatory attempts.
 Theoretical
progress in understanding the SM electroweak phase transition in the
early universe in conjunction with the experimental lower bound on the
mass of the SM Higgs boson, $m_H^{SM}>114$ GeV, leads to the
conclusion: no! In these lecture notes an introduction is
given to concepts and results which are necessary to understand
how this conclusion is reached. 
 Furthermore  I shall discuss  a few viable (so far) and rather plausible
baryogenesis scenarios  beyond the SM. 
\par
The plan  of these  notes is as follows: Section 2 contains
 some basics
of the SCM  which  are used in  the following chapters.
 Equilibrium  distributions and rough criteria
for the departure from local thermal equilibrium are  recalled. In
section 3
 a heuristic discussion of the
BAU $\eta$ is given. Then  the Sakharov
conditions for generating a baryon asymmetry within the SCM are  
discussed  and illustrated. In section
4  we review  how baryon number (B) violation
occurs in the SM and how strong B-violating SM reaction  rates are
 below and above the electroweak phase transition.
Section 5 is devoted 
to electroweak baryogenesis scenarios. The  electroweak phase 
transition is discussed,  including  results concerning its nature in the
SM which  reveal why the SM fails to explain 
the observed BAU. Nevertheless,  
electroweak baryogenesis is still a viable  scenario in  extensions
of the SM, for instance in 2-Higgs doublet and supersymmetric (SUSY)
extensions. We shall  outline this in the context of
one of the several  non-SM electroweak
baryogenesis mechanisms which  were  developed.
In section 6 we discuss  the perhaps most plausible, in any case 
most popular,  baryogenesis scenario above the electroweak phase
transition,
namely the out-of-equilibrium decay of (a) superheavy
particle(s).
After having recalled a  textbook example of 
baryogenesis in grand unified theories (GUTs), we turn to 
a  viable and attractive scenario that has found much 
attention in recent years, which is  baryogenesis through
leptogenesis caused by the decays of heavy Majorana neutrinos.  A summary
and outlook is given in section 7. 
Some formulae concerning the transformation properties of the
baryon number operator and the properties of Majorana neutrino fields
are contained in appendices A and B, respectively.
\par
Throughout these lectures the natural units of particle physics are
used
in which $\hbar = c = k_B = 1$, where $k_B$ is the Boltzmann constant.
In these units we have, for instance, that 1 GeV $\simeq$ $10^{13}K$
and  1$({\rm GeV})^{-1}\simeq 6\times 10^{-25} s$. Moreover, it is
useful to recall that the present extension of the visible universe
is characterized by the Hubble distance $H_0^{-1} \sim 10$ Gpc, where
1 pc $\simeq$ 3.2 light years.
\par 
These lectures  were intended as an introduction to the subject for
graduate students. The reader who wants to delve more deeply into these
topics
should  consult the textbook \cite{Kolb},  the  reviews 
\cite{Dolgov:1991fr,Cohen:1993nk,Rubakov:1996vz,Trodden:1998ym,Riotto:1999yt,Buchmuller:2000as}
 and, of course,  the original literature.

\section{Some Basics of Cosmology}

\subsection{The Standard Model of Cosmology}
The current understanding of the large-scale evolution of our universe
 is based on a number of observations. These include the
expansion of the universe and the  approximate isotropic and
homogeneous matter and energy distribution  on large scales.  The
Einstein field equations of general
relativity imply   
 that the metric of  space-time 
shares these  symmetry properties of
the sources of gravitation on large scales.  It is represented by the
Robertson-Walker (RW) metric which corresponds to the line element
\begin{equation}
ds^2 = dt^2 -R^2(t) \left \{ \frac{dr^2}{1-kr^2} +r^2d{\theta}^2
+ r^2\sin^2\theta \, d\theta d{\phi}^2 \right \} \: ,
\label{eq:rw} 
\end{equation}
where $(t,r,\theta,\phi)$ are the dimensionless comoving coordinates  and
$k = 0,1,-1$ for a space of vanishing, positive, or negative spatial
curvature. Cosmological data are consistent with
$k=0$ \cite{Kolb1}. The dynamical variable $R(t)$ is the cosmic scale factor and
has dimension of length.  The
matter/energy distribution on large scales may be modeled by the
stress-energy tensor  of a
perfect fluid,  $T^{\mu}_{\,\nu}=$diag$(\rho,-p,-p,-p)$,
where  $\rho(t)$ is the total
energy density of the matter and radiation in the universe and
$p(t)$ is the isotropic pressure.
\par 
The dynamical equations which determine the time-evolution of the scale
factor follow from Einstein's equations. Inserting 
the metric tensor which is encoded in  (\ref{eq:rw})
and the above form of $T_{\mu\nu}$ into these equations
 one obtains the Friedmann equation
\begin{equation}
H^2 \: \equiv \:  \left (\frac{\dot{R}}{R}\right )^2 \:  = \: 
\frac{8\pi G_N}{3}\rho -\frac{k}{R} +\frac{\Lambda}{3} \: .
\label{eq:fried} 
\end{equation}
Here $H(t)\equiv \dot{R}(t)/R(t)$ is the Hubble parameter which measures
the expansion rate of the universe at time $t$, and
$\Lambda$ denotes the cosmological ``constant'' at time t. According
to the inflationary universe scenario the $\Lambda$ term played a
crucial role at a very early epoch when vacuum energy was the dominant
form of energy in the universe, leading to an exponential increase of
the scale factor. Recent observations indicate that  today  the
largest component of the energy 
density of the universe is  some dark energy  which can also be
described by 
a non-zero cosmological  constant \cite{Kolb1}. 
The baryogenesis scenarios that we
shall discuss in these lecture notes are associated 
with a period in the evolution of the
early universe  where, supposedly, a $\Lambda$ term in the evolution equation
(\ref{eq:fried}) for $H$ can be neglected. 
\par
The  covariant conservation
of the stress tensor  $T_{\mu\nu}$ yields another important equation,
namely
\begin{equation}
d(\rho R^3) = -pd(R^3) \: .
\label{eq:1law} 
\end{equation}

\begin{figure}[ht]
\begin{center}
\includegraphics[width=9cm]{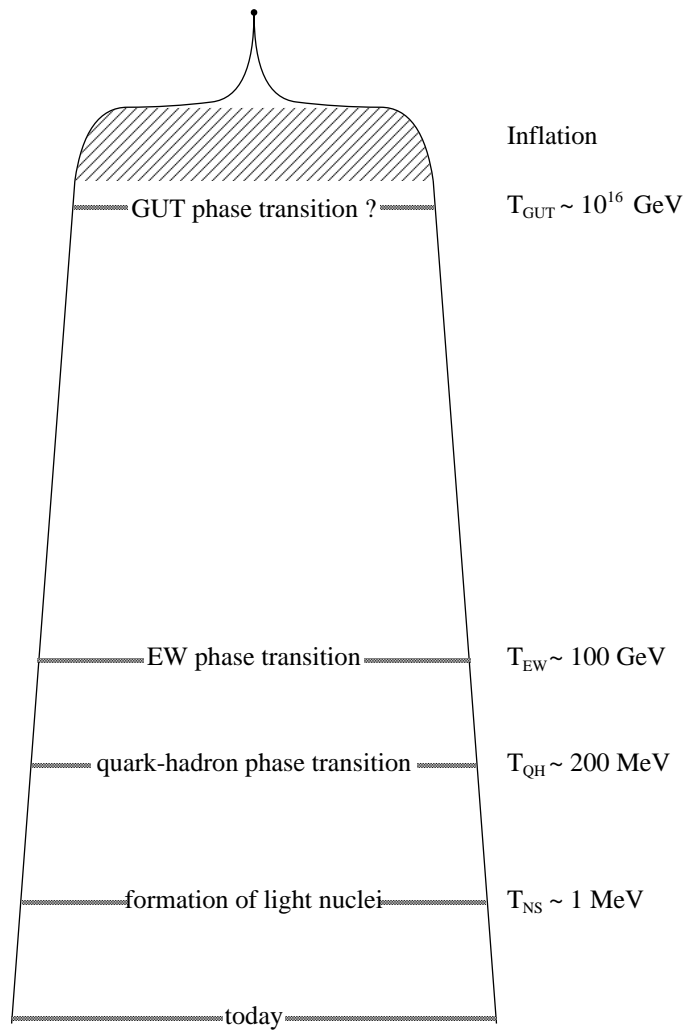} \\ 
\end{center}
\vspace*{-.5cm}
\caption[]{ 
Cartoon of the history of the universe. The 
slice of a cake, stretched  at the top, 
 illustrates the expansion of the universe as it cooled  off. Inflation may
 have ended  well below $T_{GUT}$ \cite{Riotto:1999yt}.
\protect\label{fig:p3}}
\end{figure}

This can be read as the first law of thermodynamics: the total change
of energy is equal to the work done on the universe, $dU = dA = -pdV$. 
Moreover, it turns out (see section 2.2) that the various forms of matter/energy which
determine the state of the universe during a  certain epoch
 can be described, to a good approximation, by the 
equation of state 
\begin{equation}
p = w\rho \: ,
\label{eq:estate} 
\end{equation}
 where, for instance, $w=1/3,0,-1$ if
the energy of the universe is dominated by relativistic particles
(i.e., radiation), non-relativistic particles, and vacuum energy, respectively.
\par
 Integrating (\ref{eq:1law}) with (\ref{eq:estate}) one  obtains
that the energy density evolves as $\rho \propto R^{-3(1+w)}$.
In the  radiation-dominated era, $\rho \propto R^{-4}$. Inserting this
scaling law into the Friedmann equation, 
 one finds that in this epoch the expansion rate behaves
as
\begin{equation}
H(t) \propto t^{-1} .
\label{eq:Htime} 
\end{equation}
\par
Fig. \ref{fig:p3} illustrates  the history of the early
universe, as reconstructed by the SCM and by the  SM of particle
physics.
The baryogenesis scenarios which will be discussed in sections 5 and 6 
apply to  some instant 
in the -- tiny -- time interval after inflation and before or at the
time of the electroweak phase transition. In this era, 
where the SM particles were massless,
the  energy of the universe was -- according to what is 
presently known  -- essentially due to relativistic
particles.

\subsection{Equilibrium Thermodynamics}
As was just mentioned the baryogenesis scenarios which  we shall discuss in
sections 5 and 6 apply to the era between the end of inflation and the
electroweak phase transition. During this period the universe expanded
and cooled off to temperatures $T\gtrsim T_{EW}\sim 100$ GeV. 
For most of the time during this stage 
the reaction rates of the majority of  particles 
were much faster than the expansion
rate of the cosmos. The early universe, which we view as a (dense)
plasma  of particles,  was then
to a good approximation  in thermal equilibrium. 
In several situations  it is  reasonable
to   treat this  gas as dilute and weakly interacting\footnote{This is
of course not true in general. 
The early universe   contained, in particular,
 particles that carried
unscreened non-abelian gauge charges. Such a  plasma
behaves in many ways differently
than an ideal gas.}. 
Let's  therefore recall the equilibrium
distributions of an ideal  gas.
 Because particles in the early universe 
were  created and destroyed, it is natural to  describe the gas by
means of the grand canonical ensemble. 
Consider an ensemble of a 
 relativistic particle species A.  
Its  phase space distribution or occupancy function  is given by 

\begin{equation}
f_A({\bf p}) = \frac{1}{e^{(E_A-\mu_A)/T_A} \mp 1} \: ,
\label{eq:eqdist} 
\end{equation}
where $T_A$ is the temperature, 
$\mu_A$ is the chemical potential of the species which is
associated with a  conserved charge of the ensemble, and the minus
(plus) sign refers to bosons (fermions). If different species are in 
chemical equilibrium then their chemical potentials are related. For
instance, suppose the  particle reaction  $A +B \leftrightarrow C$
takes place rapidly. Then the relation
 $\mu_A +\mu_B=\mu_C$ holds. Take the standard example $e^+ + e^-
\leftrightarrow n\gamma$. Because $\mu_{\gamma}=0$ we have $\mu_{e^+}
=-\mu_{e^-}$.
\par
From (\ref{eq:eqdist}) one obtains the number density $n_A$, the
energy density $\rho_A$, the isotropic pressure $p_A$, and the entropy
density $s_A$. Defining $d{\tilde p}\equiv  d^3p/(2\pi)^3$ we have
\begin{eqnarray}
n_A & = & g_A \int d{\tilde p} \, f_A({\bf p}) \: , \\
\rho_A & = & g_A \int d{\tilde p} \: E_A({\bf p}) f_A({\bf p}) \, , \\
p_A &  = & g_A \int d{\tilde p} \: \frac{\bf p^2}{3 E_A} f_A({\bf p}) \, , \\
s_A &  = & \frac{\rho_A + p_A}{T} \, .
\label{eq:eqfunc}
\end{eqnarray}
Here $E_A=\sqrt{{\bf p^2} +m_A^2}$, where $m_A$ is the mass of A, and
$g_A$ denotes the internal degrees of freedom of A; for instance, $g_e
= 2$ for the electron and $g_{\nu} = 1$ for a massless neutrino.
\par
In the following we need these expressions in the ultra-relativistic
($T_A>> m_A$) and nonrelativistic ($T_A <<  m_A$) limits. Integrating
eqs. (7) - (9) one obtains the well-known textbook formulae
for $n_A$, $\rho_A$, and $p_A$. 
For relativistic particles A (and $T_A >>\mu_A$)
\begin{eqnarray}
n_A &  = & a_A g_A T_A^3 \, , \\
\rho_A & = & b_A g_A T_A^4 \, , \\
p_A  & \simeq & \rho_A/3 \, , 
\label{eq:ultrar}
\end{eqnarray}
while for nonrelativistic particles the number density becomes exponentially
suppressed for decreasing temperature:
\begin{eqnarray}
n_A & =  & g_A(\frac{ m_A T_A}{2\pi})^{3/2} e^{-(m_A-\mu_A)/T_A} \, , \\
\rho_A & = & n_A m_A \, , \\
p_A  & \simeq & n_A T_A << \rho_A \,  .
\label{eq:nonr}
\end{eqnarray}
In eqs. (11), (12) $a_A$ and $b_A$ are numbers depending  on
whether A is a boson or fermion. Eqs.
(\ref{eq:ultrar}), (\ref{eq:nonr}) are the equations of
state that we used already above. \par
When considering the total energy density and pressure of all particle
species it is useful to express these quantities in terms of the 
photon  temperature T. The corresponding formulae are obtained in a
straightforward fashion by summing the respective 
contributions,  taking into account that some species A may have a thermal
distribution with a temperature $T_A \neq T$. 
When the universe was in  thermal equilibrium its
 entropy  remained
constant. Its entropy density is given by
\begin{equation}
s =\frac{S}{V} = \frac{\rho + p}{T} =\frac{2\pi^2}{45} g_{*s}T^3 \, ,
\label{eq:sdens} 
\end{equation}
where the last equality comes from the fact that $s$ is dominated
by the contributions from relativistic particles. During the
epoch we are interested in,  the factor 
$g_{*s}$ was  equal to the total number of relativistic
 degrees of freedom $g_{*}$ \cite{Kolb}. (For $T>> m_{top}$ we have
$g_{*} \simeq 106$ in the SM.) The entropy being constant implies
$s\propto R^{-3}$, hence $g_{*s}T^3 R^{3}= const$. From this we
obtain that in the radiation dominated epoch the temperature of the
universe decreased  as 
\begin{equation}
T \propto R^{-1} \, .
\label{eq:tdecr} 
\end{equation}
From these relations we can draw another important conclusion.
Consider the number $N_A$ of some particle species A. Because $N_A\equiv R^3 n_A \propto
n_A/s$
this ratio also remained  constant, in the absence of ``A number''
violation 
and/or
entropy production,  during the expansion
of the universe. Therefore in the context of
baryogenesis the relevant quantity is the baryon-to-entropy ratio
$n_B/s\equiv (n_b - n_{\bar b})/s$, where $n_b$ and  $n_{\bar b}$ denotes
the number density of baryons and antibaryons, respectively. 
The BAU $\eta \equiv n_B/n_{\gamma}$ is given in terms of this ratio
by  $\eta =1.8 g_{*s} n_B/s$.
The relativistic degrees of freedom 
 $g_{*s}$ decreased during the expansion of the
early universe. This number and, hence,  $\eta$ remained  constant 
only after the
time of $e^+e^-$ annihilation. 
From then on  
\begin{equation}
\eta \simeq 7 \frac{n_B}{s} .
\label{eq:etas} 
\end{equation}

\subsection{Departures from Thermal Equilibrium}
Departures from thermal equilibrium (DTE) were, of course, crucial for the
development of the universe to that state that we perceive today. 
Examples for DTEs include the decoupling of neutrinos, the decoupling
of the photon background radiation, and primordial nucleosynthesis. More
speculative
examples are inflation, first order phase transitions in the early
universe (see below), the decoupling of weakly
interacting massive particles, and the topic of these lectures,
baryogenesis.
 In any case the DTEs have led to the
(light) elements, to a net baryon number of the visible universe,
and to the neutrino and the microwave background.
\par
A  rough criterion for whether or not a particle
species A is in local thermal equilibrium is obtained by comparing 
reaction rate $\Gamma_A$ with the expansion rate $H$. Let 
$\sigma(A+target \to X)$ be the total cross section of the reaction(s)
of A that is (are)  crucial for keeping A in thermal
equilibrium. Then $\Gamma_A$ is given by 
\begin{equation}
\Gamma_A =\sigma(A+target \to X) n_{target} |{\bf v}| \, , 
\label{eq:rerate} 
\end{equation}
where $n_{target}$ is the target density and ${\bf v}$ is the relative
velocity. Keep in mind  that $[\Gamma_A] =(sec)^{-1}$. If
\begin{equation}
\Gamma_A \gtrsim H \, , 
\label{eq:AgH} 
\end{equation}
then the reactions involving A occur rapidly enough for A to maintain
thermal equilibrium. If
\begin{equation}
\Gamma_A < H \, , 
\label{eq:AlH} 
\end{equation}
then the ensemble of particles A will fall out of  equilibrium.
The Hubble parameter $H(t)$ which is 
 relevant for the baryogenesis scenarios to
be discussed below  is the expansion
rate during the radiation dominated epoch. It follows from  eqs.
(\ref{eq:fried}) and  (12)  that in this era
\begin{equation}
H = \sqrt{\frac{8\pi G_N}{3}}\rho = 1.66 \sqrt{g_*}\frac{T^2}{m_{Pl}} \,
,
\label{eq:Hexp} 
\end{equation}
where $m_{Pl}=1.22\times 10^{19}$GeV  denotes the Planck mass. 
\par
\begin{figure}[ht]
\begin{center}
\includegraphics[width=9cm]{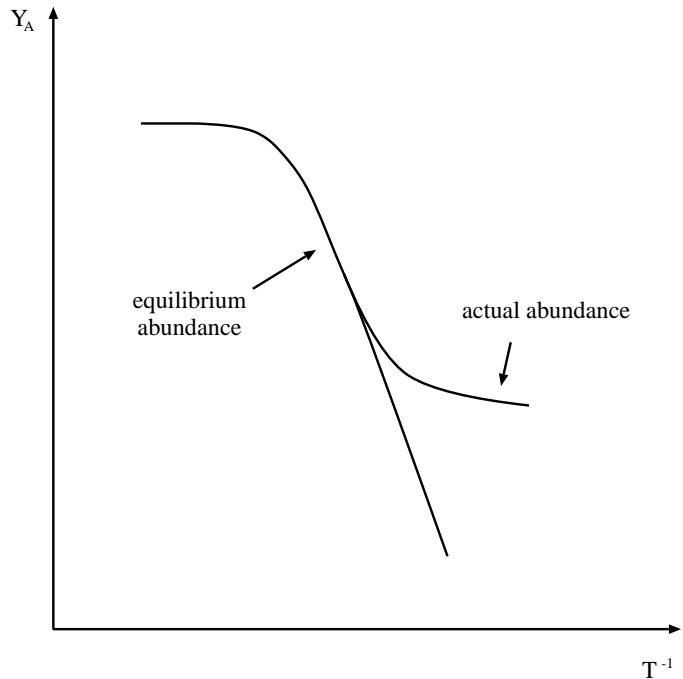} \\ 
\end{center}
\vspace*{-.5cm}
\caption[]{ 
The behaviour of $Y_A \equiv {n_A}/{s}$  as a function of decreasing
temperature
for  a massive, (non)relativistic particle species A  falling
 out of thermal equilibrium.
\protect\label{fig:p15}}
\end{figure}

Eqs. (\ref{eq:AgH}) and (\ref{eq:AlH}) constitute a useful rule of
thumb that is often quite accurate.  It is sufficient 
for the purpose of these lectures. A proper treatment involves the 
determination of the time evolution of the particle's phase space
distribution $f_A$ which is governed by the Boltzmann equation
(cf. for instance  \cite{Kolb}). Comparing the number density $n_A(t)$,
obtained from solving this equation, with the equilibrium distribution
$n_A^{eq}$ (which was discussed above for (non)relativistic particles)
one sees whether or not A has decoupled from the thermal bath. Rather
than going into details 
let us sketch in Fig. \ref{fig:p15} the behaviour of the ratio
\begin{equation}
Y_A \equiv \frac{n_A}{s} 
\label{eq:fiducial} 
\end{equation}
as a function of the decreasing temperature when
an ensemble of  massive particles A decouples from the thermal
bath. In thermal equilibrium
 $Y_A$ is constant  for $T>>m_A$. At later times,
when $T\lesssim m_A$, $Y_A\propto (m_A/T)^{3/2}\exp(-m_A/T)$ if the
reaction rate still obeys  (\ref{eq:AgH}). Thus,  if A would have remained
in thermal equilibrium until today its 
abundance would be completely negligible.  However, if $\Gamma_A$
becomes smaller than $H$, the interactions of A ``freeze out'', and
the actual abundance of A deviates from its equilibrium value at
temperature $T$. 
The larger the $A\bar A$ annihilation
cross section the smaller
the decoupling temperature and the actual abundance
$Y_A$. The
further fate of the decoupled  species depends on  whether or not A is
stable. If a (quasi)stable species A -- a weakly interacting
massive particle -- froze out  at a temperature T not
much smaller than $m_A$ then its abundance today can be significant.

\section{The Baryon Asymmetry of the Universe}

\subsection{Heuristic Considerations}
Now to the main topic, the  
matter-antimatter asymmetry of our observable universe. So far, no
primordial antimatter has been observed in the cosmos. Cosmic rays
contain a few antiprotons, $n_{\bar p}/n_p \sim 10^{-4}$, but  that number
is consistent with secondary production by protons hitting
interstellar
matter, for instance, $p +p \to 3p + {\bar p}$. Also,   
in the vicinity of the earth no
antinuclei such as  ${\bar{\rm D}}$, ${\overline{\rm He}}$ were found 
\cite{Saeki:1998rr,Alcaraz:1999ss}. 
In fact  if large, separated  domains of matter and antimatter in the
universe exist, for instance galaxies and anti-galaxies, then one
would expect annihilation at the boundaries, leading to a diffuse,
enhanced $\gamma$ ray background. However, no anomaly was observed in
such spectra. A  phenomenological analysis
 led the authors  of ref. \cite{Cohen:1998ac} to the conclusion 
that on scales larger than
100 Mpc to 1 Gpc the universe consists only of matter. While this does
not preclude
a universe with net baryon number equal to zero, no 
mechanism is known that separates 
matter from antimatter on such large scales.
\par
Thus for the visible universe
\begin{equation}
n_b - n_{\bar b} \simeq n_b \qquad 
{\Rightarrow} \qquad  \eta \simeq \frac{n_b}{n_{\gamma}} \, . 
\label{eq:nbbbb} 
\end{equation}
How is $\eta$ determined? The most direct
estimate is obtained  by  counting the number of baryons in the universe
and comparing the resulting $n_b$ with the number density of the $T =
2.7 K$ microwave photon background (CMB), 
$n_{\gamma}=2 \zeta(3)T^3/\pi^2 \simeq 420/cm^3$. 
In fact this not very precise method yields a number
 for $\eta$ that is not too far off from the
one that comes from the still most accurate  determination to date, the theory
of primordial nucleosynthesis -- a theory that is one of the triumphs
of the SCM. There the present abundances of light nuclei, $p$, ${\rm D}$,
${\rm ^3He}$, ${\rm ^4He}$, etc. are predicted in terms of the input parameter
$\eta$. Comparison with the observed  abundances yields \cite{Olive}
\begin{equation}
\eta \simeq (1.2 - 5.7) \times 10^{-10} \, . 
\label{eq:etab} 
\end{equation}
It is gratifying that the recent determination of $\eta$ from the 
CMB angular power spectrum measured by the Boomerang and MAXIMA
collaborations 
\cite{PdeBernardis} is consistent with (\ref{eq:etab}).
\par 

Can the order of magnitude of the BAU $\eta$ be
understood within the SCM, without further input? The answer is no!
The following textbook exercise shows nicely the point; namely, in order
to understand (\ref{eq:etab}) the universe must have been baryon-asymmetric
already at early times.  
The usual, plausible starting point 
of the SCM is that the big bang produces equal
numbers of quarks and antiquarks that end up
in equal numbers of
nucleons and antinucleons if 
there were no  baryon number
 violating interactions.
Let's compute the nucleon and antinucleon densities. At 
temperatures below the nucleon mass
$m_N$ we would have, as long as the (anti)nucleons
are in  thermal equilibrium,  
\begin{equation}
\frac{n_b}{n_{\gamma}} = \frac{n_{\bar b}}{n_{\gamma}} 
\simeq \left (\frac{m_N}{T}\right )^{3/2} \exp{(-m_N/T)} \, .
\label{eq:nucdist} 
\end{equation}
The freeze-out of (anti)nucleons occurs  when the $N\bar N$ annihilation rate
$\Gamma_{ann} = n_b$ $<\sigma_{ann} |{\bf v}|>$ becomes  
smaller than the expansion
rate. Using $\sigma_{ann} \sim 1/m_{\pi}^2$ and using eq. (\ref{eq:Hexp}) we find that
this happens at $T \simeq 20$ MeV. Then we have from (\ref{eq:nucdist}) that at the time
of feeze-out ${n_b}/{n_{\gamma}}= {n_{\bar b}}/{n_{\gamma}} \simeq 10^{-18}$, which
is 8 orders of magnitude below the observed value! In order to prevent $N\bar N$ annihilation
some unknown mechanism must  have operated  at $T\gtrsim 40$ MeV, 
the temperature when  ${n_b}/{n_{\gamma}}= 
{n_{\bar b}}/{n_{\gamma}} \simeq 10^{-10}$,
and separated  nucleons from antinucleons. However, the causally connected region at that
time contained only about $10^{-7}$ solar masses! Hence this separation 
mechanism were completely useless for generating our universe made of baryons.
Therefore the conclusion to be drawn from these considerations is that the universe possessed
already at early times ( $T\gtrsim 40$ MeV) an asymmetry between the number of 
baryons and antibaryons. 
\par
How does this asymmetry arise?
There might have been  some (tiny) excess of baryonic charge already at the
beginning of the big bang -- even though that does not seem to be an 
attractive idea. In any case, in the context of inflation such an
initial condition becomes  futile:  at the end of the inflationary period
any trace of such a condition had been  wiped out. 

\subsection{The Sakharov Conditions}
In the early days of the big bang model $\eta$ 
was accepted as one of the fundamental parameters of the model.
 In 1967, three years after CP violation was
discovered by the observation of the decays of $K_L \to 2\pi$,
 Sakharov pointed out in his seminal
paper \cite{Sak} that 
 a baryon asymmetry can actually  arise  dynamically 
during the evolution of the universe from an initial state
with baryon number equal to zero if
the following three conditions hold:\\
$\bullet$ baryon number (B) violation,\\
$\bullet$  C and CP violation,\\
$\bullet$ departure from thermal equilibrium (i.e., an ``arrow of time''). \\
Many models of particle physics  have
these ingredients, in combination with the SCM. The theoretical
challenge 
has been to find out
which of them support (plausible) scenarios that yield the correct order of
magnitude of the BAU. Before turning to some of these models,
let us briefly discuss the Sakharov conditions.
The first one  seems  obvious -- see, however, the remark below.
The second requirement
 is easily understood, noticing
that the baryon number operator $\hat B$ is odd both 
under C and CP (see Appendix A).
Therefore a non-zero baryon number, i.e., a non-zero expectation value
$<\hat B>$ requires that the Hamiltonian $H$ of the world 
violates C and CP. 
A formal argument for condition three is as follows:
First, recall that a system which is 
in thermal equilibrium is stationary and is described by a density operator
$\rho=\exp(-H/T)$. 
Using ${\hat B(t)}=e^{iHt}{\hat B(0)}e^{-iHt}$ we have
\begin{eqnarray*}
<\hat B(t)>_T \:= tr(e^{-H/T}e^{iHt}{\hat B(0)}e^{-iHt})= 
tr(e^{-iHt}e^{-H/T}e^{iHt}{\hat B(0)})= <{\hat B(0)}>_T \, ,
\label{eq:bthequ} 
\end{eqnarray*}
If the  Hamiltonian $H$ is $\Theta\equiv CPT$ invariant, 
${\Theta}^{-1}H\Theta = H$,
we get for the quantum mechanical equilibrium 
average of $\hat B\equiv \hat B(0)$: 
\begin{eqnarray}
<\hat B>_T \:= tr(e^{-H/T}\hat B)= tr({\Theta}^{-1}\Theta
e^{-H/T}
\hat B) \nonumber \\
= tr(e^{-H/T}\Theta\hat B{\Theta}^{-1}) = - <\hat B>_T \, ,
\label{eq:bcpt} 
\end{eqnarray}
where we used that  $\hat B$ is  odd under CPT (see Appendix A). Thus
$<\hat B>_T=0$ in thermal equilibrium. 

\begin{figure}[ht]
\begin{center}
\includegraphics[width=9cm]{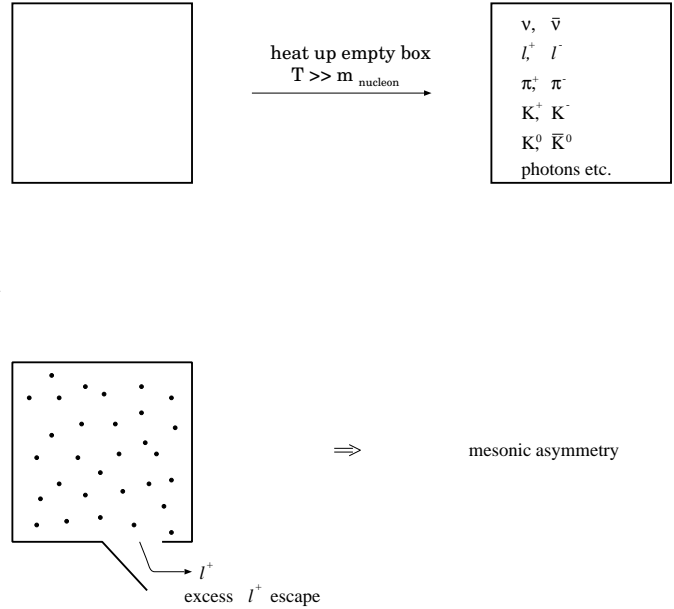} \\ 
\end{center}
\vspace*{-.5cm}
\caption[]{ 
A {\it Gedanken-Experiment} 
that illustrates two of the three Sakharov
conditions.
\protect\label{fig:p20}}
\end{figure}

\par 
How the average baryon number is kept equal to zero in thermal
equilibrium  is a bit tricky, 
as the following example shows  \cite{Dolgov:1991fr}.
Consider an ensemble of a  heavy particle species
$X$ that has 2 baryon-number
violating decay modes $X\to qq$ and $X\to \ell{\bar q}$ into quarks
and leptons. (Take $q=d$ and $\ell=e$.) Further, assume that there
is C and CP violation in these decays such that an asymmetry in the
partial decay rates of $X$ and its antiparticle $\bar X$ is induced:
\begin{equation}
\Gamma(X\to qq)\: = \: (1+\epsilon)\Gamma_0 \:, \qquad
\Gamma({\bar X}\to {\bar q}{\bar q})\: = \: (1-\epsilon)\Gamma_0 \: ,
\label{debaex}
\end{equation}
and there  will also be an asymmetry for the other channel. 
CPT invariance is supposed to
hold. Then  the total decays rates of $X$ and $\bar X$
are equal. In the decays of $X,\bar X$ a non-zero baryon number
$\Delta B$ is  generated. The ensemble is supposed to be in
thermal equilibrium. One might  be inclined to appeal to the principle
of detailed balance which would tell us that the
inverse decay $qq\to X$ is more likely than 
${\bar q}{\bar q}\to {\bar X}$,
and the temporary excess $\Delta B\neq 0$ would be erased this way.
However, this principle is based on $T$ invariance -- 
but CPT invariance implies that  this symmetry is broken 
because of CP violation.
In fact applying a CPT transformation to the
above decays, CPT invariance tells us that the inverse decays 
push  $\Delta B$ into the same direction as (\ref{debaex}):

\begin{equation}
\Gamma(qq \to X)\: = \: (1-\epsilon)\Gamma_0 \:, \qquad
\Gamma({\bar q}{\bar q} \to {\bar X})\: = \: (1+ \epsilon)\Gamma_0 \: .
\label{de1baex}
\end{equation}
The elimination of the baryon number $\Delta B$ is achieved by
the B-violating reactions $ qq \to \ell{\bar q}$, 
$ {\bar q}{\bar q} \to {\bar\ell} q$, and the CPT-transformed
reactions, where  the $X, \bar X$ resonance contributions are to be
taken out
of the scattering amplitudes.  It is the unitarity
of the S matrix which does the job of keeping $<{\hat B}>_T=0$ in
thermal equilibrium. 
\par 
The following {\it Gedanken-Experiment}, sketched in Fig. \ref{fig:p20},
illustrates two of the three Sakharov conditions
\cite{Toussaint:1978br}. 
Let's
simulate the big bang by taking an empty box and heat it up to a
temperature, say, above the nucleon mass. Pairs
of particles and antiparticles are produced that start interacting with
each other, instable particles decay, etc. 
The $K^0$ and ${\bar K}^0$ evolve in time
as coherent  superpositions of $K_L$ and $K_S$, and these states
have CP-violating decays, for instance the observed non-leptonic
modes $K_L\to\pi\pi$, and there is the 
 observed  CP-violating charge asymmetry in the semileptonic
decays $K_L\to \pi^{\mp}\ell^{\pm}\nu$  \cite{Groom:2000in}. 
When analyzing the semileptonic decays of ${K}^0$ and
${\bar K}^0$ one finds that slightly more 
$ \pi^-\ell^+\nu_{\ell}$ are produced than
$\pi^+\ell^-{\bar\nu}_{\ell}$ , by about one part in $10^3$. 
Hence, although initially there were  equal numbers of $K^0$
and  ${\bar K}^0$, their decays 
produce more $\pi^-$ than
$\pi^+$. Yet 
as long as the system is in thermal equilibrium,
CP violation in
the  reactions  including   
 $\pi^+ \ell^- \leftrightarrow \pi^+\pi^-{\nu}_{\ell}$ 
and  
 $\pi^- \ell^+ \leftrightarrow \pi^+\pi^-{\bar \nu}_{\ell}$ 
will wash out  the 
temporary excess of $\pi^-$. However, if  a thermal
instability is created by opening the box for a while, 
the excess $\ell^+$ from
neutral kaon  decay have a chance to escape. Then the inverse reactions
involving $\ell^+$ are blocked to some degree, and a mesonic asymmetry
$(N_{\pi^-}-N_{\pi^+})>0$ is generated. Of course, we haven't yet produced
the real thing, as no B-violating interactions came into play.
\par
In general,
the Sakharov conditions are sufficient but not necessary
for generating a non-zero baryon number. Each of them can be
circumvented
in principle  \cite{Dolgov:1991fr}. 
For instance, if $H$ is not CPT invariant, the argumentation
of eq. (\ref{eq:bcpt}) fails. However, such 
ideas have so far not led to a satisfactory 
explanation of (\ref{eq:etab}).
For the baryogenesis scenarios that will be discussed in 
sections 5,6 the Sakharov 
conditions are necessary ones.

\section{CP and B Violation in the Standard Model}
The standard model of particle physics combined with the SCM has, it
seems, all the ingredients for generating a baryon asymmetry.
First we recall the salient features of
 the SM at temperatures $T\simeq 0$ which apply to
present-day physics.  The observed particle spectrum
tells us that the electroweak gauge symmetry $SU(2)_L\times U(1)_Y$,
for which there is solid empirical evidence, 
cannot be a symmetry of the ground state. In the SM this spontaneous
symmetry breaking is accomplished by  a $SU(2)_L$ doublet of scalar
fields $\Phi(x)$, the Higgs field, that is assumed to have a non-zero
ground state expectation value $<0|\Phi|0> = 246$ GeV (see
below). This  classical
 field selects a direction in the internal $SU(2)_L\times
U(1)_Y$  space and hence breaks the electroweak symmetry, leaving intact
the  gauge
symmetry of electromagnetism. 
The $W$ and $Z$ bosons, quarks,  and leptons acquire their masses
by coupling to this field (which may be viewed as a
Lorentz-invariant ether). 
\par
C and CP are violated by the charged weak quark interactions
\begin{equation}
{\cal L}_{cc} \: =\: -\frac{g_w}{\sqrt 2}{\bar U}_L\gamma^{\mu}
V_{KM}D_L W_{\mu}^+ \: + \: h.c. \, .
\label{eq:lagcc} 
\end{equation}
Here $U_L=(u_L,c_L,t_L)^T,$ $ D_L=(d_L,s_L,b_L)^T,$ 
denote the left-handed quark fields ($q_L=(1-\gamma_5)q/2$), $ W_{\mu}^+$ 
is the W boson field,
 $g_w$ is the weak gauge coupling,
and $V_{KM}$ is the Cabibbo-Kobayashi-Maskawa  mixing
matrix.
CP is violated if the KM phase angle $\delta_{KM} \neq 0,\pm\pi$. By this
``mechanism'' the
CP effects observed so far in  the $K$ and
$B$ meson systems (cf., e.g., \cite{Klein,Ali,Waldi}) can be explained.
\par
There is also baryon number violation in the SM, but this is
a subtle, non-perturbative effect which is completely negligible for
particle reactions in the 
laboratories at present-day collision energies, but very significant
for the physics of the early universe. Let us outline how this effect
arises. 
From experience we know that baryon and lepton number, which are
conventionally
assigned to quarks and leptons as given in the table,  are good
quantum numbers in particle reactions in the laboratory.  
\begin{center}\renewcommand{\arraystretch}{1.5}
\begin{tabular}{c|cccc} 
 & $q$ & {$\bar q$}& $\ell$ &${\bar\ell}$  \\ \hline
B & 1/3 & -1/3& 0 & 0 \\ 
L  & 0 & 0 & 1 & -1 \\  
\end{tabular}
\end{center}
In the  SM this is explained by the circumstance that the SM Lagrangian
${\cal L}_{SM}(x)$,  with its strong-interaction (QCD) and electroweak parts,
 has a  global  $U(1)_B$ and $U(1)_L$ symmetry:
${\cal L}_{SM}$ is invariant under  the
following two sets of global  phase transformations of the quark and lepton 
fields\footnote{Possible right-handed Dirac-neutrino
degrees of
freedom are of no concern to us  here. Majorana neutrinos that
lead to violation of lepton number -- see Appendix B -- 
would be evidence for physics beyond the SM.}
$q=u,...,t;$ $\ell=e,...,\nu_{\tau}$:

\begin{eqnarray}
q(x)&  \to & e^{i\omega/3}q(x) \;, \hspace*{1cm} \ell(x)  \to  \ell(x) \:, \\
\ell(x) & \to & e^{i\lambda}\ell(x) \;, \hspace*{1.4cm} q(x)  \to  q(x) \:. 
\label{eq:u1phase} 
\end{eqnarray}
Applying Noether's theorem we obtain the associated symmetry
currents $J^B_{\mu}$ and  $J^L_{\mu}$, which  are conserved at the Born
level:
\begin{equation}
\partial^{\mu}J^B_{\mu} \:  = \: \partial^{\mu}
\sum_q \frac{1}{3}{\bar q}\gamma_{\mu} q \: = \: 0 \, , 
\label{eq:cocur1}
\end{equation}
\begin{equation}
\partial^{\mu}J^L_{\mu} \:  =  \: \partial^{\mu}
\sum_{\ell}{\bar \ell}\gamma_{\mu} \ell \: = \: 0 \, . 
\label{eq:cocur2} 
\end{equation}
(The currents are to be normal-ordered.) Thus the associated 
charge operators
\begin{equation}
{\hat B} \:  =   \: \int d^3x J^B_{0}(x) \, , 
\label{eq:bcharge} 
\end{equation}
\begin{equation}
{\hat L} \:   =  \: \int d^3x J^L_{0}(x) \,  
\label{eq:lcharge} 
\end{equation}
are time-independent. 
At the level of quantum fluctuations
 beyond the Born approximation these symmetries are, however,
 explicitly broken because eqs. (\ref{eq:cocur1}), (\ref{eq:cocur2}) no longer hold.
This is seen as follows. Decompose the vector current
\begin{equation}
{\bar f}\gamma_{\mu} f \: = \: {\bar f}_L\gamma_{\mu} f_L \: +
 \:{\bar f}_R\gamma_{\mu} f_R \, ,
\label{eq:flr} 
\end{equation}
where $f=q,\ell$, into its left- and right-handed pieces. Because
of the  clash between gauge and chiral symmetry at the quantum
level the gauge-invariant chiral currents are not conserved: in the
quantum theory the
current-divergencies suffer from  the Adler-Bell-Jackiw anomaly
\cite{Adler:gk,Bell:ts}. For
a gauge theory based on a gauge group $G$, which is a simple
Lie group  of dimension $d_G$,  
the anomaly equations  for the L- and R-chiral 
currents ${\bar f}_L\gamma_{\mu} f_L$ and 
${\bar f}_R\gamma_{\mu} f_R$ read
\begin{eqnarray}
\partial^{\mu} {\bar f}_L\gamma_{\mu} f_L \: & = & \: -c_L\frac {g^2}{32
  \pi^2} F^a_{\mu\nu}{\tilde F}^{a\mu\nu} \, , \\
\partial^{\mu} {\bar f}_R\gamma_{\mu} f_R \: & = & \: +c_R\frac{g^2}{32
  \pi^2} F^a_{\mu\nu}{\tilde F}^{a\mu\nu} \, , 
\label{eq:abjan} 
\end{eqnarray}
where $F^{a\mu\nu}$ is the (non)abelian field strength
tensor ($a=1,..., d_G$) and ${\tilde F}^{a\mu\nu} =\epsilon^{\mu\nu\alpha\beta}
F^a_{\alpha\beta}/2$ is the dual tensor,\footnote{We use the
convention $\epsilon_{0123}=+1$.}
  $g$ denotes the gauge
coupling, and the constants $c_L, c_R$ depend on the representation
which  the $f_L$ and $f_R$ form. 
Let us apply (\ref{eq:flr}) - (\ref{eq:abjan}) to the above baryon and lepton
number currents of the SM where the gauge group is $SU(3)_c\times SU(2)_L\times
U(1)_Y$. Because gluons couple  to right-handed and
left-handed quark currents with the same strength, we have $c_L^{QCD} =
c_R^{QCD}$. Therefore $J^B_{\mu}$ has no QCD anomaly. However,  the
weak gauge bosons $W_\mu^{a}, a=1,2,3,$ 
couple only to left-handed quarks and leptons,  while the weak
hypercharge boson couples to $f_L$ and $f_R$ with different strength.
Hence $c^W_R=0$ and  $c_L^Y \neq
c_R^Y$.  Putting everything together one obtains 
\begin{equation}
\partial^{\mu}J^B_{\mu} \:  = \: 
\partial^{\mu}J^L_{\mu} \:  =  \: \frac{n_F}{32\pi^2}
(-g_w^2 W^a_{\mu\nu}{\tilde W}^{a\mu\nu} + g'^2 
B_{\mu\nu}{\tilde B}^{\mu\nu}) \, , 
\label{eq:waekan} 
\end{equation}
where 
$W^a_{\mu\nu}$ and $B_{\mu\nu}$ denote the $SU(2)_L$ and $U(1)_Y$
field strength tensors, respectively, $g'$ is the  $U(1)_Y$ gauge
coupling,
and $n_F=3$ is the number of generations. 
\par
\begin{figure}[ht]
\begin{center}
\includegraphics[width=9cm]{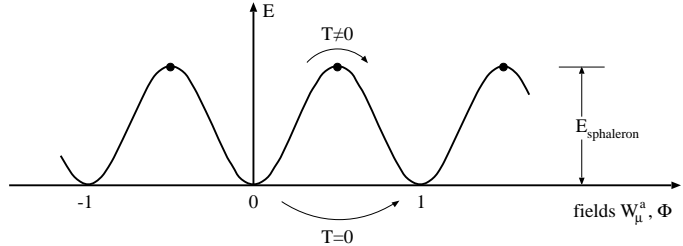} \\ 
\end{center}
\vspace*{-.5cm}
\caption[]{ 
The periodic vacuum structure of the standard electroweak theory. 
The direction in field space has been chosen as described in the text.
The schematic diagram  shows the energy of static gauge and  Highs field 
configurations
$W^a_{\mu}({\bf x}), \Phi({\bf x})$. The integers
are the  Chern-Simons
number $N_{CS}$ of the respective zero-energy field configuration.
\protect\label{fig:p27}}
\end{figure}

Eq. (\ref{eq:waekan}) implies that $\partial^{\mu}(J^B_{\mu}-
J^L_{\mu})=0 $. Thus the difference
of the baryonic and leptonic charge operators ${\hat B}-{\hat L}$ remains
time-independent also at the quantum level and therefore the quantum
number \\ \\
\centerline{B - L is conserved in the SM.}
\\ \\
How does B+L number violation come about?
We note that the right hand side of
eq. (\ref{eq:waekan}) can also be written as the divergence
of  a current $K^{\mu}$:
\begin{equation}
r.h.s. \; {\rm of} \;  (\ref{eq:waekan}) 
\: = \: n_F \partial_{\mu} K^{\mu} \, ,
\label{eq:dicscurr} 
\end{equation}
where
\begin{equation}
K^{\mu} \: = \; - \frac{g_w^2}{32\pi^2}2\epsilon^{\mu\nu\alpha\beta}
W^a_{\nu}(\partial_{\alpha}W^a_{\beta} + \frac{g_w}{3}\epsilon^{abc}
W^b_{\alpha}
W^c_{\beta}) + \frac{g'^2}{32\pi^2}\epsilon^{\mu\nu\alpha\beta}
B_{\nu}B_{\alpha\beta} \: .
\label{eq:cscurr} 
\end{equation}
Let's integrate  eq. (\ref{eq:waekan}), using (\ref{eq:dicscurr}),
  over space-time. Using Gau{\ss}'s
law we convert these integrals into integrals over a surface at
infinity.
Let's first do the surface integral for the right-hand side of  
(\ref{eq:waekan}). For hypercharge gauge fields $B_{\mu}$ with
acceptable behaviour  at  infinity, that is,  vanishing field strength
$B_{\alpha\beta}$, the abelian part of $K_\mu$ makes no contribution
to this integral. For the non-abelian gauge fields $W^a_{\nu}$
vanishing field strength implies that $2\epsilon_{\mu\nu\alpha\beta}
\partial^{\alpha}W^{a\beta}= -g_w\epsilon_{\mu\nu\alpha\beta}\epsilon^{abc}
W^{b\alpha}
W^{c\beta}$ at infinity. Using this we obtain
\begin{equation}
\int d^4x \, \partial^{\mu}K_{\mu} \: = \;  
\frac{g_w^3}{96\pi^2}\int_{\partial V_4} dn^{\mu}\, 
\epsilon_{\mu\nu\alpha\beta}\epsilon^{abc}W^{a\nu}W^{b\alpha}
W^{c\beta}  \: .
\label{eq:czsumf} 
\end{equation}
Now we choose the surface $\partial V_4$ to be a large 
 cylinder with top and bottom surfaces
at $t_f$ and $t_i$, respectively, and let the volume of the cylinder
tend to infinity.  Because  $\partial_{\mu}K^{\mu}$ is
gauge-invariant, we  may choose a special gauge. Choose 
the temporal gauge condition,
$W^a_0=0$. Then there is no contribution from the integral over the
coat
of the cylinder and we obtain 
\begin{equation}
\int d^3xdt \, \partial_{\mu}K^{\mu} \: = \;  
N_{CS}(t_f) - N_{CS}(t_i) \equiv \Delta N_{CS}\, ,
\label{eq:csumf} 
\end{equation}
where
\begin{equation}
N_{CS}(t) \: =\: \frac{g_w^3}{96\pi^2}\int   d^3x\, 
\epsilon_{ijk}\epsilon^{abc}W^{ai}W^{bj}
W^{ck} 
\label{eq:csnum} 
\end{equation}
is the Chern-Simons number. This integral assigns a topological
``charge'' to a  classical gauge field. Actually $N_{CS}$ is not gauge
invariant
but  $\Delta N_{CS}$ is.  A nonabelian gauge theory
like weak-interaction  $SU(2)_L$ is topologically non-trivial,
 which is reflected by
the fact that it has an infinite number of ground states whose vacuum 
gauge field configurations have different topological charges
 $\Delta N_{CS} = 0, \pm 1,\pm 2, \ldots$  Imagine the set of gauge and
Higgs fields
and consider the energy functional $E[field]$ that forms a hypersurface  
over this infinite-dimensional space.
The ground states with different
topological charge are separated by a potential barrier. In 
Fig. \ref{fig:p27}
a one-dimensional slice through this hypersurface is drawn. The
direction in field space has been chosen such that the  classical path
from one ground state to another goes over a pass of minimal height. 

Finally we perform the 
 integral over the left-hand sides of
(\ref{eq:waekan}) and  get the result
\begin{equation}
\Delta{\hat B} \: =\: \Delta{\hat L } \: =\: n_F \Delta N_{CS} \, ,
\label{eq:blopv} 
\end{equation}
with  $\Delta{\hat Q}\equiv {\hat Q}(t_f) -{\hat Q}(t_i),$ $Q=B,L$.
Eq. (\ref{eq:blopv}) is to be interpreted as follows. As long as we
consider small gauge field quantum fluctuations around the
perturbative vacuum configuration $W^a_{\mu}=0$ the right-hand side 
of (\ref{eq:blopv}) is  zero, and $B$ and $L$ number remain
conserved. This is the case in perturbation theory to arbitrary order
where B- and L-violating processes have zero amplitudes. However, large gauge
fields $W^a_{\mu}\sim 1/g_w$ with nonzero topological charge  
$\Delta N_{CS} =  \pm 1,\pm 2, ...$ exist. As discovered by `t Hooft 
\cite{'tHooft:up}
they can induce transitions at the quantum level between
fermionic states $|i,t_i>$ and $|f,t_f>$ with  baryon and 
lepton numbers that differ  according to the rule (\ref{eq:blopv}): 
\begin{equation}
\Delta{ B} \: =\: \Delta{ L } \: =\: n_F \Delta N_{CS} \, .
\label{eq:blzahl} 
\end{equation}
This selection rule tells us that
 $B$ and $L$ must change by at least 3 units.\footnote{Notice that, even after taking these non-perturbative effects into account,  the SM still predicts
the proton to be stable.}
A closer inspection of the global $U(1)$
symmetries and associated currents shows that, in situations where
fermion masses  can be neglected,  the selection rule 
can be refined: there is a change in quantum numbers by the same amount for
 every generation. Thus, e.g., 
 $ \Delta{ L_e }= \Delta{ L_{\mu} }= \Delta{ L_{\tau} }
=\Delta{ B}/3 =\Delta N_{CS}$.
\par
\begin{figure}[ht]
\begin{center}
\includegraphics[width=9cm]{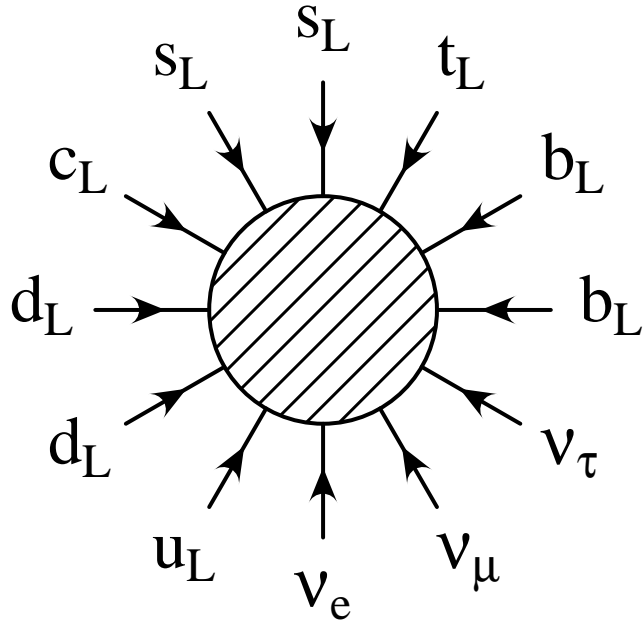} \\ 
\end{center}
\vspace*{-.5cm}
\caption[]{ 
An example of a  (B+L)-violating standard model amplitude. 
The arrows indicate the flow
of the fermionic quantum numbers.
\protect\label{fig:p26}}
\end{figure}

The dominant B- and L-violating transitions are between  states
 $|i,t_i>$ and  $|f,t_f>$  where  
$|\Delta{ B}|  = |\Delta{ L }|$ changes by 3 units. 
At temperature $T=0$, 
transitions with
$|\Delta{ B}|  = |\Delta{ L }|=3$ are  induced by the (anti)instanton
\cite{Belavin:fg}, a
gauge field which connects two vacuum configurations
whose topological charge differ by $\pm 1$.
When put into the temporal gauge $W^a_0=0$ then the
instanton field $W^a_i({\bf x},t)$ approaches, for instance,
  $W^a_{i}=0$ at $t_i\to -\infty$ and  
a topologically non-trivial vacuum configuration with $N_{CS}=1$
at $t_f\to +\infty$, as indicated in 
 Fig. \ref{fig:p27}. The corresponding
  amplitudes $<f,t_f|i,t_i>$
involve 9 left-handed quarks (right-handed $\bar q$) -- where each generation 
participates with 3 different color states -- and 3 left-handed leptons
 (right-handed $\bar \ell$), one of each generation. One of the
possible amplitudes is depicted in Fig. \ref{fig:p26}. 
Hence we have, for instance, the anti-instanton induced
reaction with $\Delta{ B}  = \Delta{ L }= -3$:
\begin{equation}
u + d \to {\bar d} + 2{\bar s} + {\bar c} + 2{\bar b} +{\bar t} +
{\bar\nu_e} + {\bar\nu_{\mu}} + {\bar\nu_{\tau}} \, .
\label{eq:blreact} 
\end{equation}

What is the probability for such a  transition to occur? It is clear
from Fig. \ref{fig:p27} that it corresponds to a  tunneling
process. Thus it must be exponentially suppressed. The classic
computation of `t Hooft \cite{'tHooft:up,'tHooft:fv}
implies, for energies $E_{c.m.}(ud)\lesssim {\cal O}(1\, TeV)$,
 a  cross-section 
\begin{equation}
\sigma_{\Bsla + \Lsla} \: \propto \: e^{-4\pi/\alpha_w} \: \sim \:
  10^{-164} \,,
\label{eq:bltho} 
\end{equation}
where $\alpha_w=g_w^2/4\pi\simeq 1/30$.
\par
When  the standard model is coupled to
a heat bath of temperature $T$, the situation changes. 
As was first shown in \cite{Kuzmin:1985mm} (see also \cite{Arnold:1987mh}),
at very high temperatures $T\gtrsim T_{EW}\sim 100$ GeV the B- and L-violating
processes in the SM are fast enough to play a significant role in baryogenesis.
In order to understand this we have again a look  at  Fig. \ref{fig:p27}.
The ground states with different $N_{CS}$ are separated by a potential barrier
of minimal height
\begin{equation}
E_{sph}(T) \: = \: \frac{4\pi}{g_w} v_T f(\frac{\lambda}{g_w}) \: ,
\label{esph} 
\end{equation}
where  $v_T \equiv <0|\Phi|0>_T$ is the vacuum expectation value (VEV)
of the SM Higgs doublet field $\Phi(x)$ at temperature T. 
At $T=0$ we
have  $v_{T=0}= 246$ GeV. The parameter $f$ varies between $1.6 <f<2.7$ 
depending on the value of the Higgs self-coupling $\lambda$, i.e., on the
value of the SM Higgs mass. This yields $E_{sph}(T=0) \simeq 8 - 13$ TeV.
The subscript ``sph'' refers to the sphaleron, a gauge and Higgs field 
configuration of Chern-Simons number 1/2 (+ integer) 
which is an (unstable) solution of the classical field 
equations of the SM gauge-Higgs sector \cite{Manton:1983nd,Klinkhamer:1984di}. 
These kind of field configurations (their locations 
are indicated by the dots in Fig. \ref{fig:p27}) 
lie on the respective minimum 
energy path from
one ground state to another with different Chern-Simons number. 
 Fig. \ref{fig:p27} suggests that the rate of fermion-number non-conserving 
transitions will be proportional
to the Boltzmann factor $\exp(-E_{sph}(T)/T)$ as long as the energy of the
thermal excitations is smaller than that of the barrier, while unsuppressed
transitions will occur above that barrier. 
\par
At this point we recall that the electroweak (EW) $SU(2)_L\times U(1)_Y$
 gauge symmetry was unbroken at high temperatures, that is, in the early 
universe.  The critical temperature $T_{EW}$ where -- running backwards 
in time --  the transition from the broken phase with Higgs 
VEV $v_T \neq 0$ to the
symmetric phase with  $v_T =  0$ occurs is, in the SM, about  100 GeV. (A 
discussion of this transition will be given in the next section.)
Hence the B- and L-violating transition rates of the SM 
will no longer be exponentially suppressed  above this temperature.
Detailed investigations have led to the following results: \\
$\bullet$ In the phase where the EW gauge is broken, i.e., $T<T_{EW}\sim 100$ 
GeV, the sphaleron-induced ${\Bsla + \Lsla}$ transition rate per
volume $V$
is given by (see, e.g., \cite{Rubakov:1996vz,Moore:1998sw})
 
\begin{equation}
 \frac{\Gamma^{sph}_{\Bsla + \Lsla}}{V} 
= \kappa_1 \left (\frac{m_W}{\alpha_wT}\right )^3 m_W^4 \exp(-E_{sph}(T)/T) \, ,
\label{eq:sprbel} 
\end{equation}
where $m_W(T)= g_w v_T/2$ is the temperature-dependent mass of the W boson
and $\kappa_1$ is a dimensionless constant. \\
$\bullet$ The calculation of the transition rate in the unbroken phase
is very difficult. On dimensional grounds we
expect this rate per volume to be
 proportional to $T^4$. Recent investigations \cite{Bodeker:1999gx,Moore:2000mx} yield
 for $T>T_{EW}\sim 100$ GeV:
\begin{equation}
\frac{\Gamma^{sph}_{\Bsla+\Lsla}}{V} 
= \kappa_2 \, \alpha_w^5 T^4 \, ,
\label{eq:sprabo} 
\end{equation}
with $\kappa_2 \sim 21$. 
\par 
By comparing $\Gamma^{sph}_{\Bsla + \Lsla}$ above $T_{EW}$ with the expansion 
rate
$H$  given in  (\ref{eq:Hexp}), we can assess whether the (B+L)-violating
SM reactions, which conserve B-L, are fast enough to keep up with
the expansion of the early universe in the radiation dominated epoch.  
From the requirement $\Gamma^{sph}_{\Bsla + \Lsla} >> H$ one obtains
that these processes are in thermal equilibrium for temperatures
\begin{equation}
T_{EW}\; \sim \; 100 \: {\rm GeV}\; <\; T\; \; \lesssim \; 10^{12}\: {\rm GeV} \: .
\label{eq:blcons} 
\end{equation}
This result provides an important constraint on any baryogenesis mechanism
which  operates above $T_{EW}$. If the B- and L-violating 
interactions involved 
in this mechanism conserve B-L, then any excess of baryon and lepton number
generated above $T_{EW}$ will be washed out 
by the  B- and L-nonconserving SM sphaleron-induced reactions. 
Hence baryogenesis scenarios 
above  $T_{EW}$ must be based on particle physics models that violate also
B-L. Examples will be discussed in section 6. 

\section{Electroweak Baryogenesis}
We haven't discussed yet which phenomenon could possibly
provide the  third Sakharov ingredient, the departure from thermal equilibrium,  
if one attempts to explain the baryon asymmetry  within the  SM of particle physics.
A little thought reveals that
a  baryogenesis scenario based on the SM requires  that  the
thermal instability must come from the electroweak phase transition. 
First of all,  the expansion rate of the universe at temperatures,
say, $T\lesssim 10^{12}$ GeV  is
too slow for causing a departure from local thermal equilibrium:
the reaction rates of most of the SM particles, which are typically
of the order of
$\Gamma\sim \alpha_w^2 T$ or larger, are much larger than the expansion rate
(\ref{eq:Hexp}), even for extremely high temperatures. 
Further, the SM charged
weak quark current interactions lead to  CP-violating effects only because, apart
from a non-trivial KM phase,  the u- and d-type quarks have non-degenerate masses
(see eq. (\ref{JJ})  below). These masses are generated at the EW transition,
while  all SM 
particles are massless above $T_{EW}$. 
If $\Delta B\neq 0$ was  created  at the  EW transition it would be
-- if the phase change was strongly first order --
frozen in during the later evolution of the universe, as the B- and L-violating 
reactions below $T_{EW}$ would be strongly suppressed 
(see eq. (\ref{eq:sprbel}) 
and below). However, the  investigations of refs.
\cite{Buchmuller:1995sf,Fodor:1994sj,Kajantie:1995kf}
 have shown that the EW transition in 
the SM fails to provide the required thermal instability.
\par
\begin{figure}[ht]
\begin{center}
\includegraphics[width=9cm]{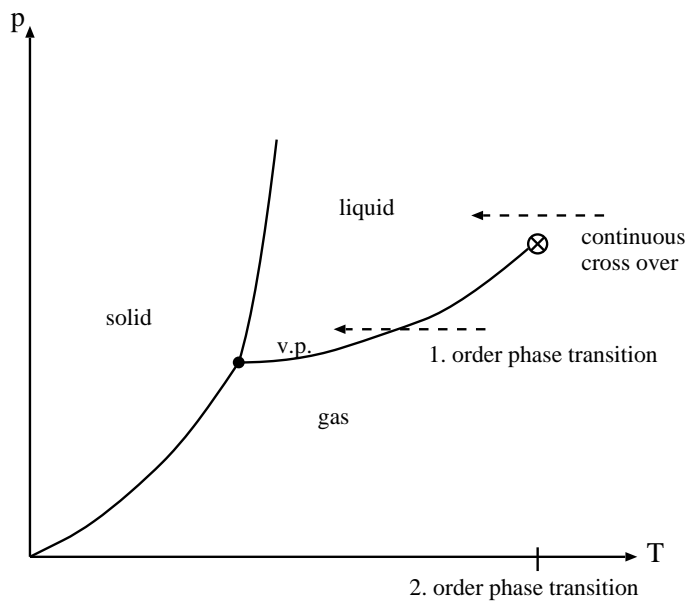} \\ 
\end{center}
\vspace*{-.5cm}
\caption[]{ 
The phase diagram of water.
\protect\label{fig:p38}}
\end{figure}

\begin{figure}[ht]
\begin{center}
\includegraphics[width=9cm]{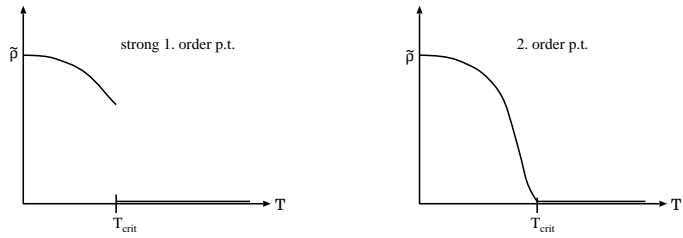} \\ 
\end{center}
\vspace*{-.5cm}
\caption[]{ 
The behaviour of the order parameter  ${\tilde\rho}$ in the case of
a  strong first
order and  a second order transition.
\protect\label{fig:p38b}}
\end{figure}

 
Before reviewing the results on the nature of the
EW transition in the SM let us recall 
some basic concepts about phase transitions. Consider Fig. \ref{fig:p38}
where the pressure versus temperature phase diagram of water is sketched. We
concentrate on the  vapor $\leftrightarrow$ liquid transition. 
The curve to the  right of 
the triple point  is  the so-called vapor-pressure curve. 
For values of $p, T$ along this line there is  a coexistence of the liquid and 
gaseous phases. A change of the parameters 
across this curve leads to a first
order phase transition which  becomes weaker along the curve. The endpoint 
corresponds to a second order transition.  Beyond that point there is a smooth 
cross-over from the gaseous to the liquid phase and vice versa. 
The nature of a phase transition can be  characterized by an  order 
parameter appropriate to the system.  For the 
vapor-liquid transition the order parameter is the
difference in the densities of water in the liquid and gaseous phase,
${\tilde\rho}= \rho_{liquid}-\rho_{vapour}$. In the case of a strong first
order phase transition the order 
parameter has a strong discontinuity at the
critical temperature $T_c$ where the transition occurs:
in the example at hand ${\tilde\rho}$ is very small in the vapor phase but
it makes a sizeable jump at $T_c$ because of the coexistence of both phases --
see Fig. \ref{fig:p38b}. That's what we need in a successful  
EW baryogenesis scenario!
In case of a second order phase transition the order parameter changes also
rapidly in the vicinity of $T_c$, but the change is continuous. In the 
cross-over region of the phase diagram the   continuous
change of ${\tilde\rho}$ as a function of $T$ is less pronounced.

\newpage

\begin{figure}[ht]
\begin{center}
\includegraphics[width=7cm]{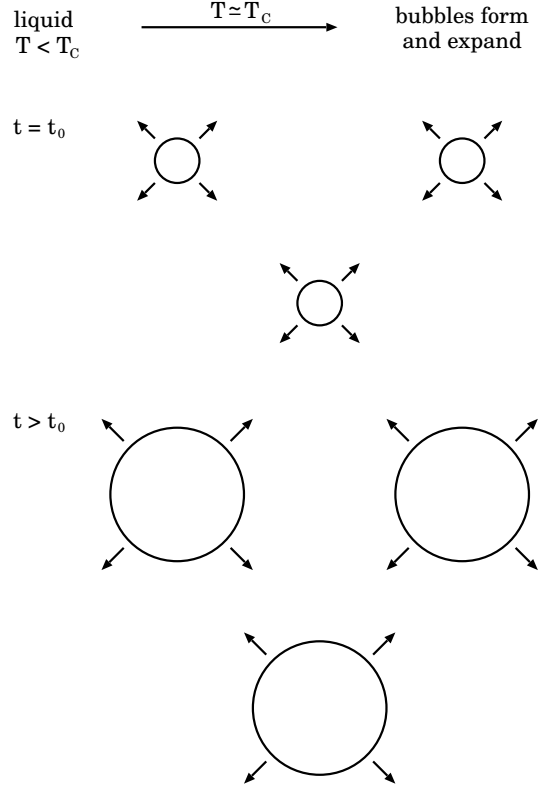} \\ 
\end{center}
\vspace*{-.5cm}
\caption[]{ 
Dynamics of a first-order liquid-vapor phase transition: Formation 
and expansion
of vapor bubbles.
\protect\label{fig:p39}}
\end{figure}

\par 
So far to the statics of phase transitions. As to their dynamics, 
we know from
experience how the  first-order liquid-vapor transition evolves in time. 
Heating up water, vapor bubbles start to nucleate
slightly below $T=T_c$ within the liquid. They
expand and finally percolate above $T_c$. This is illustrated in 
Fig \ref{fig:p39}. Drawing the analogy to the early
universe  we should, of course,  rather consider the cooling of vapor and
its transition to a liquid through the formation of droplets.

A standard  theoretical method to determine the 
nature of a  phase transition
in a classical system, like the  vapor$\leftrightarrow$liquid or
paramagnetic$\leftrightarrow$ferromagnetic transition is as
follows. Let  ${\cal H}={\cal H}(s)$ be the classical Hamiltonian of
the system, 
where $s({\bf x})$ is a (multi-component) classical field. 
In the case of water $s({\bf x})$ is the local density, while for
a magnetic material $\vec{s}({\bf x})$ denotes the three-component
local magnetization. From the computation
of  the partition function $Z$ we obtain the
Helmholtz free energy $F= -T\ln Z$ from which  the thermodynamic
functions of interest can be derived. In particular we can 
compute the order parameter
 $s_{av}=<\sum_{\bf x}s({\bf
  x})>_T$ and study its behaviour as a function of temperature. 
\par
The investigation  of the static thermodynamic properties of gauge field
theories
proceeds along the same lines. In the case of the standard electroweak
theory the role of the order parameter is played by the VEV of the
$SU(2)_L$ Higgs doublet field $\Phi$. This becomes obvious when we
recall the following. Experiments tell us  that 
the $SU(2)_L\times U(1)_Y$ gauge symmetry is broken at $T=0.$ For the SM
this means that the mass parameter in the Higgs potential must be tuned
such that there is a non-zero Higgs VEV. On the other hand
it was shown a long time ago \cite{Kirzhnits:ut}
that at temperatures significantly larger
than, say, the W boson mass the Higgs VEV is zero and the  
$SU(2)_L\times U(1)_Y$ gauge symmetry is restored. (This will be
shown below.)
Hence during the evolution of the early
universe the Higgs field must  have condensed at some $T=T_c$. The
order of this phase transition is deduced from the behaviour of the
Higgs VEV (and other thermodynamic quantities) around $T_c$. 
\par
Let's couple the standard electroweak theory to a heat bath of
temperature $T$. The free energy $F= -T\ln Z$ is obtained from
the Euclidean functional integral 
\begin{equation}
F(J,T) \:=\: -T\ln{\left[ \int_{\beta}{\cal D}[{\rm fields}]
\exp(- \int_{\beta} dx ({\cal L}_{EW}+J\cdot\Phi))\right ]} \: ,
\label{eq:helmfe} 
\end{equation}
where ${\cal L}_{EW}={\cal L}_{EW}(\Phi, W_{\mu}^a,B_{\mu},q,\ell)$
denotes the Euclidian version of the electroweak SM Lagrangian, 
$J$ is an auxiliary external field, $\beta=1/T$,  
\begin{equation}
\int_{\beta} dx =\int^{\beta}_0 d\tau \int_V d^3x \: ,
\label{eq:hint} 
\end{equation}
and the subscript $\beta$ on the functional integral indicates that
the bosonic (fermionic) fields satisfy (anti)periodic boundary
conditions at $\tau =0$ and $\tau =\beta$. From the free energy
 density $F(J,T)/V$ the effective potential $V_{eff}(\phi,T)$ is
 obtained by a Legendre transformation, where $\phi=\partial
 F/\partial J|_{J=0}$ is the expectation value of the Higgs
 doublet field, $\phi=<\Phi>_T$.  (Actually in order to 
compare with numerical lattice calculations it is useful
to employ a gauge-invariant order parameter.)
 Recall that  the
effective potential $V_{eff}(\phi,T)$ is the energy density of the
system in that state $|a>_T$ in which the expectation value
$<a|\Phi|a>_T$ takes the value $\phi$. Hence by computing the
stationary point(s), $\partial V_{eff}(\phi,T)/\partial\phi = 0$,
the ground-state expectation value(s)  $\phi$=$<0|\Phi|0>_T$
of $\Phi$ at a given
temperature $T$ are determined.  If at some $T=T_c$ two minima 
 are found then this signals two coexisting phases and a first
order phase transition. 

\subsection{Why the SM fails}
Let us now discuss the effective potential of the SM. At $T=0$ the
tree-level
effective potential is just the classical Higgs potential
$V_{tree}=-\mu^2(\Phi^\dagger\Phi)+\lambda(\Phi^\dagger\Phi)^2$.
Choosing the unitary gauge, $\Phi^{unitary}=(0,\phi/\sqrt{2})$ with
$\phi\geq 0$ we
have
\begin{equation}
V_{tree}(\phi) \: = \: -\frac{\mu^2}{2}\phi^2 +\frac{\lambda}{4}
\phi^4 \: , 
\label{eq:vhiggs} 
\end{equation}
where $\lambda>0$ and, by assumption, $\mu^2>0$ in order that the
Higgs
field is non-zero in the 
state of minimal energy: 
$\phi_0 \equiv <0|\Phi|0>_{T=0}\equiv v_{T=0}/\sqrt{2}
=\sqrt{\mu^2/\lambda}$, and  $v_{T=0}$ is fixed by, e.g., the experimental
value of the W boson
mass to $v_{T=0}=246$ GeV.
The mass of the SM Higgs boson is given by
\begin{equation}
m_H \: =  v_{T=0}\sqrt{2\lambda} +{\rm quantum}\; {\rm  corrections}\: .
\label{eq:higgsma} 
\end{equation}
The experiments at LEP2 have established the lower bound
$m_H>114$ GeV \cite{LEPEW}. Hence the SM Higgs self-coupling $\lambda > 0.33$.
\par
\begin{figure}[ht]

\begin{center}
\includegraphics[width=9cm]{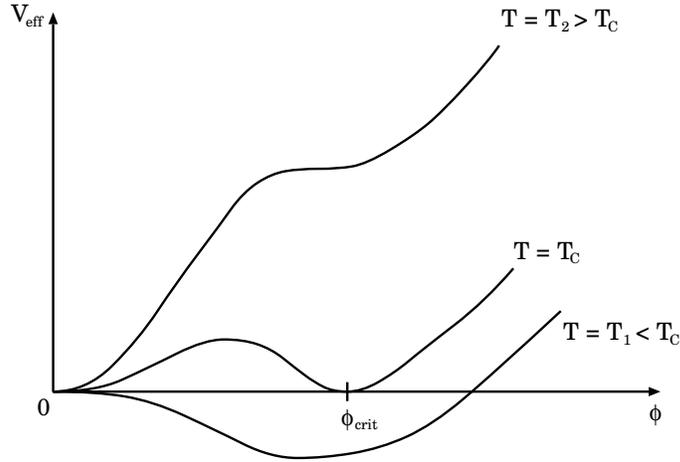} \\ 
\end{center}
\vspace*{-.5cm}
\caption[]{ 
Behaviour of  $V_{eff}$ in the case of a first order
phase transition.
\protect\label{fig:p42}}
\end{figure}

At $T\neq 0$ the SM effective potential is computed at the quantum level
as outlined above. Because the gauge  coupling $g'$ and the Yukawa
couplings of quarks and leptons $f\neq t$ ($t$ denotes the top quark) 
to the Higgs doublet
$\Phi$ are small,  
the contributions of the hypercharge gauge boson  and of $f\neq t$
may be neglected. This is usually done in the literature. Let us first
discuss, for illustration, the
effective potential computed to one-loop approximation
for the now obsolete  case of  a  very light Higgs boson. 
 For  high temperatures $V_{eff}$ is  given by 
\begin{equation}
V_{eff}(\phi,T ) \: = \frac{1}{2} a(T^2-T_1^2)\phi^2 -\frac{1}{3}
bT \phi^3 + \frac{1}{4}\lambda  \phi^4 \: ,
\label{eq:veff1l} 
\end{equation}
where 
\begin{equation}
a \:=\: \frac{3}{16}g_w^2 +(\frac{1}{2}+\frac{m_t^2}{m_H^2})\lambda\;
, \; b\:=\: 9\frac{g_w^3}{32\pi} \; , \; T_1\:=\;
\frac{m_H}{2\sqrt{a}} \; ,
\label{eq:vefpar} 
\end{equation}
and $m_t$ is the mass of the top quark. The term cubic in $\phi$ is
due to fluctuations at $T\neq 0$. If the Higgs boson was light the
quartic term would be  small. Inspecting eq. (\ref{eq:veff1l}) we recover
the result quoted above that at high temperatures the Higgs field is
zero in the ground state. When the temperature  is lowered  we
find that at $T_c= T_1/\sqrt{1-2b^2/(9a\lambda)}>T_1$ a first order
phase transition occurs: 
the effective potential $V_{eff}$ has two energetically 
degenerate minima: one at
$\phi=0$ and the other at 
\begin{equation}
v_{T_c} \: \equiv \: \phi_{crit} \: = \: \frac{2b}{3\lambda} T_c \: , 
\label{eq:vcrit} 
\end{equation}
separated by an energy barrier, see Fig. \ref{fig:p42}.
At $T_c$ the free energy of the symmetric and of the broken phase are
equal;
however, the universe remains for a while in the symmetric phase because
of the energy barrier. As the universe expands and cools down further, 
bubbles filled with the
Higgs condensate start to  nucleate at some temperature below $T_c$.
 These bubbles  become larger by releasing latent heat, 
percolate, and eventually fill the whole volume at  $T=T_1$. 
Bubble nucleation and expansion  are
 non-equilibrium phenomena which are difficult to
 compute.

\par
\begin{figure}[ht]
\begin{center}
\includegraphics[width=9cm]{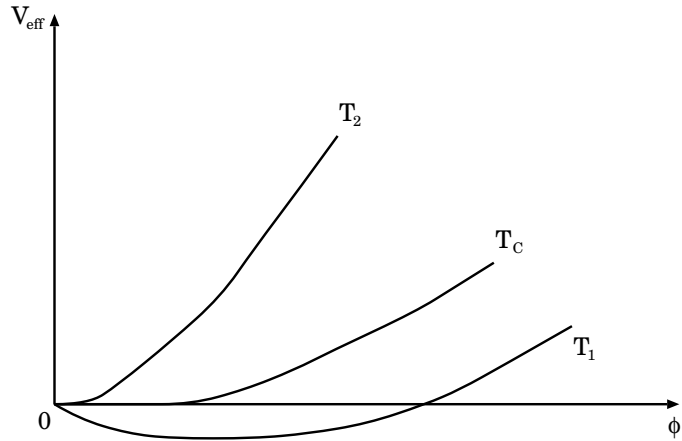} \\ 
\end{center}
\vspace*{-.5cm}
\caption[]{ 
Behaviour of  $V_{eff}$ in the case of a second order
phase transition.
\protect\label{fig:p42b}}
\end{figure}

 
Fig. \ref{fig:p42b} shows the behaviour
of $V_{eff}(\phi,T )$ in the case of a second order phase
transition. In this case there are no energetically
degenerate  minima separated by a barrier at $T=T_c$, i.e.,  
no bubble nucleation and expansion.
The Higgs field gradually condenses uniformly at $T\lesssim T_c$ and
grows to its present value as the system cools off.

The value of the critical temperature depends on the parameters of
the respective model and is obtained by
detailed computations (see the references given below).
 Nevertheless, we may use the above formula for
$T_c$ for a crude estimate and obtain $T_c \sim$ 70 GeV for $m_H=100$
GeV. (For a more precise value, see below.) With 
eqs. (\ref{eq:Htime}) and (\ref{eq:Hexp}) we then estimate that the
EW phase transition took place at a time $t_{EW} \sim 5\times 10^{-11}$
s after the big bang.
This implies that the causal domain, the
diameter of which 
is given by $d_H(t)=2t$ in the radiation-dominated
era, was then of the order of a few centimeters. 
\par
Back to baryogenesis. It should be clear now why a strong
first order EW phase transition is required. 
In this case the time scale associated with the nucleation and
expansion of Higgs bubbles is comparable with the time scales of the
particle reactions. This 
causes  a departure from thermal equilibrium. 
How is this to be quantified? Let's consider one  of the bubbles with $v_T\neq
0$ which, after expansion and percolation,  eventually become our world. 
The bubble  must get filled with more
quarks than antiquarks such that $n_B/s \sim 10^{10}$  and this ratio
remains conserved. This means that baryogenesis has to take place
outside of the bubble  while the sphaleron-induced (B+L)-violating
reactions must be strongly suppressed within the bubble.
In order that the sphaleron rate, which in the broken phase
is given by eq. (\ref{eq:sprbel}), $\Gamma^{sph}_{\Bsla + \Lsla}
\propto \exp{(-4\pi f v_T/g_w T)}$, is practically switched off, 
the  order parameter must jump  at $T_c$, from  $\phi=0$
in the symmetric phase to a value $v_{T_c}$ 
in the broken phase such that 
\begin{equation}
 \frac{v_{T_c}}{T_c} \: \gtrsim \: 1  \: .
\label{eq:vcrit1} 
\end{equation}
This is the
condition for a  first order transition to be strong.
\par
In view of the experimental lower bound
$m_H^{SM}>114$ GeV,  the  formulae
(\ref{eq:veff1l}),  (\ref{eq:vefpar}) for $V_{eff}$ which are
valid only for a very light Higgs boson  no longer
apply. Nevertheless,
eq. (\ref{eq:vcrit}) shows that the discontinuity gets weaker when the
Higgs mass is increased. 
The strength  of the electroweak  phase transition has been studied
for the  SM $SU(2)$ gauge-Higgs model as a function of the Higgs boson
mass with analytical methods \cite{Buchmuller:1995sf}, and numerically
with 4-dimensional \cite{Fodor:1994sj} and 3-dimensional
\cite{Kajantie:1995kf}  
lattice methods. These results quantify the
qualitative features discussed above: the strength
of the  phase transition changes from strongly first order
($m_H\lesssim$ 40 GeV) to weakly first order as the Higgs mass is
increased, ending at $m_H\simeq $ 73 GeV 
\cite{Buchmuller:1996pp,Rummukainen:1998as,Fodor:1999at} where the phase
transition
is second order (cf. the liquid-vapor transition discussed above).
The corresponding critical temperature is $T_c \simeq$ 110 GeV 
\cite{Laine:2000xu}. 
For larger values of $m_H$ there is a smooth cross-over between the
symmetric
and the broken phase.
\par
Thus the result of the LEP2 experiments, $m^{SM}_H > $114 GeV, leads
to  the following conclusion: 
 if the SM Higgs mechanism provides  the correct
description of electroweak symmetry breaking then the EW phase
transition in the early universe does not provide the thermal
instability required for baryogenesis. The B-violating sphaleron
 processes are only adiabatically switched off 
during  the transition from $T>T_c$
to $T<T_c$; they are still thermal for   $T\lesssim T_c$.
Thus the standard model of particle physics cannot explain the BAU
$\eta$ -- irrespective of the role  that SM  CP  violation
may play in this game.

\subsection{EW Phase Transition in SM Extensions}
Of course, whether or not the SM Higgs field  or some other mechanism 
provides  the correct description of EW symmetry breaking remains to
be
clarified. In fact,  this is the most important unsolved problem of present-day
particle physics. Future collider experiments hope to resolve
this issue. On the theoretical side, a number of
extensions and  alternatives to the
SM Higgs  mechanism  have been discussed for quite
some time. One may distinguish between models which  postulate
elementary Higgs fields (i.e., the associated spin-zero particles
have pointlike couplings up to some high energy scale $E\gg 100$ GeV)
which trigger the breakdown of $SU(2)_L\times U(1)_Y$, and others
which assume that it is caused by the Bose condensation of (new) heavy
fermion-antifermion pairs. The dynamics of the
symmetry breaking sector of these models can change the order of the
EW phase transition, as compared with the SM. Let's briefly
discuss results for some models that belong to the first class.
The presently most popular extensions of the SM are supersymmetric
(SUSY) extensions, in particular the minimal supersymmetric standard
model (MSSM), the Higgs  sector of which  contains two Higgs
doublets.
Although the requirement of SUSY breaking to be soft does not allow
for independent quartic couplings in the Higgs potential
$V(\Phi_1,\Phi_2)$,  the number of parameters of the scalar sector of
this model is larger than that of the SM and a first order 
transition can be arranged.\footnote{In models with 2 Higgs doublets
the EW phase transition typically proceeds in 2
stages, because  the 2 neutral 
scalar fields condense, in general, at 2 different temperatures
\cite{Land:1992sm,Hammerschmitt:fn}.} Investigations of
$V_{eff}$ at $T\neq 0$ show that there is a region
 in the MSSM parameter space which  allows for a
sufficiently strong first order EW phase transition (see, for instance,
 the reviews \cite{Laine:2000xu,Cline:2000fh} and references therein).
The condition for this is that  the
mass of the scalar partner $\tilde t_R$ of the
right-handed top quark $t_R$ must be sufficiently light and
the mass of $\tilde t_L$ must be sufficiently heavy. 
An upper bound on the mass of the lightest neutral Higgs boson $H_1$ of the
model obtains from the requirement that the mass of  $\tilde t_L$
should not be unnaturally large. In summary, the MSSM predicts a
sufficiently strong 1st order EW phase transition if 
\begin{equation}
m_{H_1} \: \lesssim \: 105 \: - \: 115 \: {\rm GeV} \, , \hspace*{1cm}
 m_{\tilde t_R} \: \lesssim \: 170 \: {\rm GeV} \, \: .
\label{eq:susup} 
\end{equation}
\par 
In the next-to-minimal SUSY model which contains an additional gauge
singlet Higgs field a strong first order transition can be arranged
quite easily \cite{Huber:2000mg}. 
\par
Non-supersymmetric SM extensions may be, in general, less motivated than SUSY
models,  but several
of these models are, nevertheless, worth to be studied  as
they predict interesting phenomena. 
For illustrative purposes we  mention here only the class of
2 Higgs doublet  models (2HDM) where the field content of the SM is extended by
an additional Higgs doublet, leading to a physical particle spectrum
which includes  3 neutral and one charged Higgs particle.
The general, renormalizable and  $SU(2)_L\times U(1)_Y$ 
invariant Higgs potential $V(\Phi_1,\Phi_2)$ contains a large number
of unknown  parameters. 
Therefore, it is not surprising 
that in these models, too, 
the requirement of a strong 1st order EW transition can be arranged
quite easily as  studies  of the finite-temperature effective potential 
show (see, for instance, \cite{Cline:1996mg}). No tight
upper bound on the mass of the lightest Higgs boson obtains.

\subsection{CP Violation in SM Extensions}
Another aspect of SM extensions, namely
non-standard CP violation,  is also essential for baryogenesis
scenarios. SM extensions as those mentioned above
involve, in particular, an extended non-gauge
sector; that is to say, a richer set of Yukawa and Higgs-boson
 self-interactions
than in the SM. It is these interactions that break, in general,
CP invariance. Thus, in SM extensions  
additional sources of 
CP violation besides the KM phase are usually present. 
We shall confine ourselves to 2 examples. (For a review,
see for instance \cite{Bernreuther:1998ju}.)

\subsubsection{Higgs sector CPV}
An interesting possibility is CP violation (CPV) by an extended Higgs sector
which can occur already in the 2-Higgs doublet  extensions of the SM. Consider
the class of 2HDM which are constructed such that flavour-changing 
neutral (pseudo)scalar currents are absent at tree level. The
appropriate\footnote{Neutral flavor conservation is enforced
by imposing a discrete symmetry, say, $\Phi_2
\to - \Phi_2$,  on ${\cal L}$
that may be softly broken by $V(\Phi_1,\Phi_2)$.} 
 $SU(2)_L\times U(1)_Y$ 
invariant tree-level Higgs potential $V(\Phi_1,\Phi_2)$ of these models may be
represented in the following way: 
\begin{eqnarray}
V_{tree}(\Phi_1,\Phi_2) \: &= & \: \lambda_1(2\Phi_1^\dagger\Phi_1 - v_1^2)^2
 +  \lambda_2(2\Phi^{\dagger}_2\Phi_2 - v_2^2)^2 \nonumber \\
&&  + \,  \lambda_3[(2\Phi^{\dagger}_1\Phi_1 - v_1^2)  +
 (2\Phi^\dagger_2\Phi_2-v_2^2)] \nonumber \\
&&  + \, \lambda_4[(\Phi_1^\dagger\Phi_1)(\Phi_2^\dagger\Phi_2) -
 (\Phi_1^\dagger\Phi_2)(\Phi_2^\dagger\Phi_1)] \nonumber \\
&& + \,  \lambda_5 [2{\rm Re}(\Phi_1^\dagger\Phi_2) - v_1v_2\cos\xi]^2
\nonumber \\
&& + \,  \lambda_6 [2{\rm Im}(\Phi_1^\dagger\Phi_2) - v_1v_2\sin\xi]^2 \: ,
\label{eq:v2hdm} 
\end{eqnarray}
where $\lambda_i, v_1, v_2$ and $\xi$ are real parameters and the
parameterization of $V_{tree}$ is chosen such that the Higgs fields
have non-zero VEVs in the state of minimal energy. 
\par
Performing 
a CP transformation,
\begin{equation}
 \Phi_{1,2}(\mbox{\bf x},t) \quad \stackrel{CP}{\longrightarrow} 
\quad e^{i\alpha_{1,2}}\, \Phi_{1,2}^\dagger (-\mbox{\bf x},t) \: ,
\label{VCP}
\end{equation}
we see that $H_{V}=\int d^3x \, V_{tree}(\Phi_1,\Phi_2)$ is CP-noninvariant if
 $\xi  \neq 0$.
Notice that it is unnatural to assume $\xi = 0$. Even if this was so at
tree level, the
non-zero KM phase $\delta_{KM}$, which is needed to explain 
the observed CPV in $K$ and $B$ meson decays, 
would induce a non-zero $\xi$
through radiative corrections. 
\par
From eq. (\ref{eq:v2hdm}) we read off that
 at  zero temperature the  neutral components of the Higgs
doublet fields  have,  in the electric charge
conserving
ground state, 
the expectation values 
\begin{equation}
<0|\phi^0_1|0> \: =  \: v_1e^{\xi_1}/\sqrt 2 \, , \qquad
<0|\phi^0_2|0> \:
 =  \: v_2
e^{i\xi_2}/\sqrt 2 \, , 
\label{VEV}
\end{equation}
where $v=\sqrt{v_1^2+v_2^2} = 246$ GeV, and  $\xi_2-\xi_1=\xi$ is the
physical CPV phase. 
\par The spectrum of physical Higgs boson states of  the 
two-doublet models consists of a charged Higgs
boson and its antiparticle, $H^\pm$, and three neutral states. As far 
as CPV is concerned, $H^\pm$ carries
the KM phase. This particle  affects the (CPV) phenomenology of 
flavor-changing $|\Delta F| = 2$  neutral meson mixing
and $|\Delta F| = 1$ weak decays of mesons and baryons.
( Experimental data on $b \to s + \gamma$ imply that this particle must
be quite heavy, 
$m_{H^+} > $ 210 GeV.) 
\par
Let's briefly discuss 
some implications of Higgs sector CPV
for present-day physics. If $\xi$ were zero, the set of neutral 
Higgs boson states would consist of two scalar (CP=1) and one pseudoscalar
(CP= --1) state. If $\xi\neq 0$ these states mix.
As a consequence the 3  mass eigenstates, $|\varphi_{1,2,3}>$,
no longer have
a definite CP parity. That is, they couple 
both to scalar and to pseudoscalar quark and lepton currents.
In terms of Weyl fields the corresponding Lagrangian reads
\begin{equation}
{\cal L}_{\varphi} = -  \sum_\psi
 c_\psi \frac{m_\psi}{v} {\bar \psi_L}\psi_R  \varphi \: +\: h.c. \: .
\label{Yphi}
\end{equation}
The sum over the Higgs fields $i=1,2,3$ is implicit,
 $\psi$ denotes a quark or lepton
field, 
$m_\psi$ is the mass of the associated
particle, and the dimensionless reduced Yukawa couplings  $c_\psi =
a_\psi +i b_\psi$ 
($a_\psi, b_\psi$ real) depend on the parameters
of the Higgs potential and on the type of model.
\par 
The Yukawa interaction (\ref{Yphi})
leads to CPV in $flavour$-$diagonal$ reactions 
for quarks and for leptons $\psi$. The induced CP effects are 
proportional
to some power $(m_\psi)^p$.  
For example, consider the reaction $ \psi{\bar \psi} \to
\psi{\bar \psi}$.  The exchange  of a $\varphi$  boson at tree level 
induces an effective CPV interaction
of the form $({\bar \psi}\psi)({\bar \psi}i\gamma_5 \psi)$ with a 
coupling strength proportional
to $m_{\psi}^2/m^2_{\varphi}$.
The search for non-zero electric dipole moments
(EDM) of the electron and the neutron
has traditionally been a sensitive experimental method
to trace non-SM CP violation \cite{Gould}. 
If  a light $\varphi$ boson exists ($m_{\varphi}\sim 100$ GeV)
and the CPV phase $\xi$ is of order 1 the Yukawa interaction
(\ref{Yphi})
can induce electron and neutron EDMs of the same order of magnitude as
their present experimental upper bounds.
\par  
What happens at the EW phase transition in the early universe?
We assume that the parameters of the 2HDM are such that the transition
is strongly first order. Moreover, in order to simplify the
discussion we assume that the passage from the symmetric to the broken
phase occurs in one step, at some temperature $T_c$. Somewhat below
$T_c$ bubbles filled with Higgs fields start to nucleate and
expand. That is, the Higgs VEVs are space and time dependent. Let's
consider, for simplicity,  only  one of the bubbles
and assume its expansion   to be spherically symmetric. When the
bubble
has grown to some finite size  we can use the following
one-dimensional description. Consider  the rest frame of the
bubble wall. The wall is taken to be planar and the expansion of the
bubble is taken along the  $z$ axis. The wall, i.e., the phase boundary
 has some finite thickness
$l_{wall}$, extending from $z=0$ to $z=z_0$. 
The symmetric phase lies to the right of this boundary, $z>z_0$ while
the broken phase lies to the left, $z<0$. Thus the neutral
Higgs fields have VEVs whose magnitudes and phases vary with $z$:
\begin{equation}
<0|\phi^0_1|0>_T \: = \:
\frac{\rho_1(z)}{\sqrt 2}e^{i\theta(z)} \, , 
\qquad <0|\phi^0_2|0>_T \: = \:  \frac{\rho_2(z)}{\sqrt 2}  e^{i\omega(z)} \, .
\label{TVEV}
\end{equation}
In the symmetric phase, $z\gg z_0$, both VEVs vanish, whereas in the
broken phase the VEVs should be close to their zero temperature
values: 
\begin{equation}
\rho_i(z)  \: \simeq  \: v_i \, , \qquad  \theta(z) \: \simeq  \:
\xi_1\, , \qquad
\omega(z)
 \: \simeq  \:\xi_2 \, ,
\label{ascpp}
\end{equation}
if $z \ll 0$. The variation of the moduli and phases 
with $z$ can be determined by
solving the field equations of motion that involve the
finite-temperature effective potential of the model.  
\par
As to the couplings of the Higgs fields to fermions, we assume here
and in the following subsection, for definiteness, that all quarks and
leptons couple to $\Phi_1$ only. Then the Yukawa coupling of a 
  quark or lepton field $\psi = q, \ell$ to the neutral Higgs field
is
given by 
\begin{eqnarray}
{\cal L}_1 \:&  = & \: - h_{\psi}{\bar\psi}_L\psi_R\phi^0_1 \: + \:
h.c. \nonumber \\
\:&  = & \: - m_{\psi}(z){\bar\psi}_L\psi_R \: - \:   
m_{\psi}^*(z){\bar\psi}_R\psi_L
\: + \; \ldots \: , 
\label{lnphi}
\end{eqnarray}
where 
\begin{equation}
m_{\psi}(z) =  h_{\psi} \frac{\rho_1(z)}{\sqrt 2}e^{i\theta(z)} 
\label{cpmass}
\end{equation}
is a complex-valued mass and the ellipses in (\ref{lnphi}) indicate
the
coupling of the quantum field, i.e., the coupling of a neutral Higgs
particle to $\psi$. Thus the interaction of a fermion field  $\psi(x)$ with 
the CP-violating Higgs bubble, treated as an external, classical
background field,  is summarized
by the Lagrangian
\begin{equation}
{\cal L}_{\psi} \: =\: {\bar\psi}_Li\gamma^\mu\partial_\mu \psi_L
+ {\bar\psi}_R i\gamma^\mu\partial_\mu \psi_R
- m_{\psi}(x){\bar\psi}_L\psi_R \: - \:   
m_{\psi}^*(x){\bar\psi}_R\psi_L \: . 
\label{cphibu}
\end{equation}
In section 5.4 we shall also use the  plasma frame which is implicitly defined by requiring
the form of the particle distributions to be the thermal ones. 
In this frame the
Higgs VEVs are space- and time-dependent. The wall expands  with a
velocity $v_{wall}$. 
The interaction (\ref{cphibu}) is CP-violating because
$x$-dependent  phase $\theta(x)$ of $m(x)$. Obviously,
the field 
$\theta(x)$   cannot be
removed from ${\cal L}_{\psi}$ by redefining the fields $\psi_{L,R}(x)$.
 We shall investigate its
consequences for baryogenesis in the next subsection. 

\subsubsection{CP Violation in the MSSM}
In the minimal supersymmetric extension (MSSM) of the Standard Model 
\cite{Giudice} CP-violating phases
can appear, apart from the complex Yukawa 
interactions of the quarks yielding a non-zero
KM phase $\delta_{KM}$, in the so-called $\mu$ term in the
superpotential (i), and in  soft  supersymmetry breaking terms (ii) - (iv).
The requirement of gauge invariance and hermiticity of the Lagrangian
allows for the following new sources of CP violation:\\
i) A complex mass parameter 
$\mu_c\equiv \mu\exp(i\varphi_{\mu})$,  $\mu$ real, describing the
mixing of the two Higgs chiral superfields in the superpotential. \\
2) A  complex squared mass parameter $m^2_{12}$ describing the mixing
of the two Higgs doublets\footnote{In order 
to facilitate the comparison
with the non-supersymmetric  models, the non-SUSY
convention for the Higgs doublets is employed here; i.e., 
the same hypercharge assignment is
made
for both  $SU(2)$ Higgs doublets,  $\Phi_i = (\phi^+_i, \phi^0_i)^T$, (i=1,2)
.}  and contributes  to the Higgs potential
\begin{equation}
V(\Phi_1,\Phi_1) \: \supset \;  \mu_c \Phi^\dagger_1\cdot\Phi_2 + {\rm
  h.c.} \: ,
\label{bterm}
\end{equation} 
iii) Complex
Majorana masses $M_i$ in the gaugino mass terms ($\epsilon \equiv i\sigma_2$),
\begin{equation}
-\sum_i M_i(\lambda^T_i\epsilon \lambda_i)/2 + {\rm h.c.}, 
\label{gaug}
\end{equation}
where $i=1,2,3$ refers to the $U(1)_Y$, $SU(2)_L$ gauginos, and gluinos,
respectively. A standard assumption is that the $M_i$ have a common phase. \\
iv) Complex trilinear scalar couplings
of the
scalar quarks and scalar leptons, respectively,
 to the  Higgs doublets $\Phi_1,\Phi_2$.  
These couplings form 
 complex 3$\times$3 matrices ${A}_\psi$ in generation space.
Motivated by supergravity models it is often assumed 
that the matrices $A_\psi$  are proportional to the
Yukawa coupling matrices $h_\psi$:
\begin{equation}
{A}_\psi = A{h}_\psi, \qquad \psi  = u,d,\ell ,
\label{Aff}
\end{equation}
where $A$ is a complex mass parameter. \\
Thus the parameter set $\mu_c, m^2_{12}, M_i,$ and $A$ involves 4 complex
phases.  Exploiting two  (softly
broken) global $U(1)$ symmetries of the MSSM Lagrangian, 
two of these phases can be removed by 
re-phasing of the fields.
A common choice, we we shall also use, is a phase convention for the
fields such that the gaugino masses $M_i$ and the mass parameter
$m^2_{12}$ are real. Then 
the observable CP phases in the MSSM (besides the KM phase)
are
$\varphi_{\mu}=arg(\mu_c)$ and $\varphi_{A}=arg(A)$.
 The experimental upper bounds on the
electric dipole moments $d_e$, $d_n$ of the
electron and  the neutron put,  however, rather 
 tight constraints on  these CP phases,
in particular on $\varphi_{\mu}$. 
Even if there are correlations between 
these phases such that there are cancellations among the
contributions to $d_e$ and to $d_n$,  Ref. \cite{Bartl:1999bc} finds 
(see also \cite{Ibrahim:1997gj,Brhlik:1998zn}) that 
 $\varphi_{\mu}$ is
constrained by the data to be smaller than $\mid\varphi_{\mu}\mid
\lesssim $0.03. A way out of this constraint
would be  heavy  
first and second generation  sleptons and squarks 
with masses of order 1 TeV.
\par
What about Higgs sector CPV? In the MSSM 
the tree-level Higgs potential $V_{tree}$ is CP-invariant. 
Supersymmetry does not allow for
independent  quartic couplings in $V_{tree}$. They 
are proportional to linear combinations of the $SU(2)_L$ and $U(1)_Y$
gauge 
couplings squared. 
At one-loop order the interactions of the Higgs fields $\Phi_{1,2}$
with charginos, neutralinos, (s)tops, etc.  generate  
 quartic Higgs self-interactions of the form 
\begin{equation}
V_{eff} \: \supset \:  \lambda_1 (\Phi^{\dagger}_1 \Phi_2)^2 +
\lambda_2 (\Phi^{\dagger}_1 \Phi_2)(\Phi^{\dagger}_1 \Phi_1) 
 + \lambda_3 (\Phi^{\dagger}_1 \Phi_2)(\Phi^{\dagger}_2 \Phi_2)
+ {\rm h.c.} \:  , 
\label{VLee}
\end{equation}
in the effective
 potential. The CP phases $\varphi_{\mu}$ and $\varphi_{A}$ induce
 complex
$\lambda_{1,2,3}$. Thus,  explicit  CP violation in the Higgs sector occurs
 at the quantum level which leads to Yukawa interactions of the
 neutral Higgs bosons being of the form (\ref{Yphi}). 
\par
In the context of baryogenesis a potentially
more interesting possibility is
spontaneous CP violation at high temperatures
$T\lesssim T_{EW}$. This kind of CP
violation could not be traced any more in the laboratory! 
Ref.  \cite{Comelli:pb} pointed out that,
irrespective of whether or not  
$\varphi_{\mu}$ and $\varphi_{A}$ are sizeable,  the MSSM
effective potential receives,  at high temperatures  $T\lesssim
T_{EW}$, quite large one-loop 
corrections of
the form 
(\ref{VLee}).  
As a consequence,  the
neutral Higgs fields  can  develop  complex VEVs of the form 
(\ref{TVEV}) with a large CP-odd classical
field.   This  would  signify spontaneous CPV
 at finite temperatures, even if 
$\varphi_{\mu}$ and $\varphi_{A}$ would be  very small or even zero. 
However, ref. \cite{Huber:1999sa} finds that 
experimental constraints on the parameters of the MSSM and the
requirement of the phase transition to be strongly first order preclude
this possibility in the case of the MSSM. 
\par
Let's now come to those  CP-violating
interactions of the MSSM which are of
relevance at the  EW phase transition
and involve $\varphi_{\mu}$ and $\varphi_{A}$ 
at the tree level. 
As discussed above, there is a small, phenomenologically
acceptable range of light Higgs and light stop mass
parameters which allows for a strong first order
transition. The Higgs VEVs are of the form 
\begin{equation}
<0|\phi^0_1|0>_T \: = \: {\rho_1(z)} \, , 
\qquad <0|\phi^0_2|0>_T \: =\:  {\rho_2(z)}  \, ,
\label{suVEV}
\end{equation}
where $\rho_i$ are real and for convenience,
a  normalization convention different from the one in (\ref{TVEV}) is
 used here.
\par
These VEVs determine the interaction of  the bubble
wall with  those MSSM particles 
that couple to the Higgs fields already at the classical level. 
Inspecting where the CP-violating phases  $\varphi_{\mu}$ and
$\varphi_{A}$
are located in ${\cal L}_{MSSM}$
(we use the convention
of  the gaugino masses $M_i$ being real) it becomes clear that 
the relevant interactions of the classical Higgs background fields
are those with charginos,
neutralinos, and sfermions, in particular top squarks.
Contrary to the case of the 2HDM discussed above the interactions of
quarks and leptons  with a bubble wall do  not -- at the
classical
level -- violate CP invariance  if (\ref{suVEV}) applies.
\par   
Inserting (\ref{suVEV}) into the respective terms of the MSSM
Lagrangian we obtain the Lagrangians  describing the particle
propagation in the presence of a Higgs bubble
\cite{Huet:1995sh,Cline:2000nw,Carena:2000id}. For the charged
gauginos and Higgsinos in the gauge eigenstate basis we get 

\begin{eqnarray}
{\mathcal L}_c & = &{\chi}_R^\dagger \sigma_\mu\partial^\mu\chi_R +
{\chi}_L^\dagger{\overline\sigma}_\mu\partial^\mu\chi_L \nonumber\\
& & + 
{\chi}_R^\dagger {\mathcal M}_c \chi_L + {\chi}_L^\dagger
{\mathcal M}_c^{\dagger} \chi_R
\label{chacp}
\end{eqnarray}
where $\sigma^\mu=(I,\sigma_i)$, ${\overline\sigma}^\mu=(I,-\sigma_i)$,
and we have put 
\begin{equation}
\chi_R^\dagger =
 (
{\widetilde W^+} \, , {\widetilde{H}_2^+}  ) 
\, ,
\qquad
\chi_L =  (
{\widetilde{W}^-}\, , {\widetilde{H}_1^-} )^T \: ,
\end{equation}
where ${\widetilde{W}}(x)$, ${\widetilde{H}_{1,2}}(x)$
are 2-component Weyl fields for
the charged gauginos and Higgsinos, respectively.
The chargino mass matrix is given by
\begin{equation}
\mathcal{M}_c(x)=\left(
\begin{array}{cc}
M_2 & g_w\rho_2(x) \\
g_w\rho_1(x) & \mu_c
\end{array}
\right) \: ,
\label{ccmat}
\end{equation}
where $\mu_c$ is the complex Higgsino mass parameter defined above.
\par
For the scalar stop fields  $\tilde t_R(x)$, $\tilde t_L(x)$
we obtain in the gauge eigenstate basis

\begin{equation}
{\mathcal L}_{\tilde t}= (\partial_\mu {\tilde t_L}^\dagger)\partial^\mu
{\tilde t_L} +
(\partial_\mu {\tilde t_R}^\dagger)\partial^\mu {\tilde t_R} 
- \left(
\begin{array}{ll}
{\tilde t_L}^\dagger\, , & {\tilde t_R}^\dagger
\end{array}
\right)
{\cal M}_{\tilde t}\left(
\begin{array}{l}{\tilde t_L}\\ 
{\tilde t_R}
\end{array}\right) \, ,
\label{stopcp}
\end{equation}
with 
\begin{equation}
{\cal M}_{\tilde t}(x)=\left(
\begin{array}{cc}
m_L^2+h_t^2\,H_2^2(z) & h_t\left( A_t \rho_2(x)-\mu^*_c \rho_1(x)\right) \\
h_t\left( A_t^* \rho_2(x)-\mu_c \rho_1(x)\right) & m_R^2+h_t^2\,\rho_2^2(x)
\end{array}
\right)\ ,
\label{stcpm}
\end{equation}
where $m_{L,R}^2$ are SUSY breaking squared mass parameters,
$h_t$ is the top-quark Yukawa coupling, and $A_t$ is the left-right
stop mixing parameter. 
\par
In the mass matrices (\ref{ccmat}) and (\ref{stcpm})
the CP-violating phases combine with the spatially varying VEVs and
will give rise to $x$-dependent CP-violating phases when  the mass
matrices are diagonalized,
analogously to the case of the 2HDM above. 
This causes CP-violating particle currents which we shall discuss
further in the next subsection.

\subsection{Electroweak Baryogenesis}
\begin{figure}[ht]
\begin{center}
\includegraphics[width=9cm]{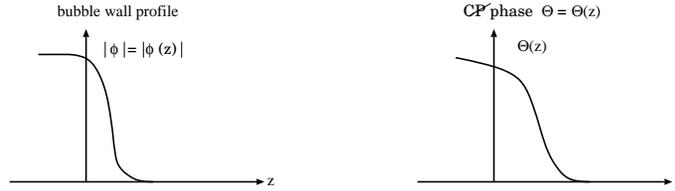} \\ 
\end{center}
\vspace*{-.5cm}
\caption[]{ 
Sketch of the variation of the modulus and the CP-violating  phase angle 
of a non-SM Higgs VEV, in the wall frame,  at the boundary
between the broken and the symmetric phase.
\protect\label{fig:p46b}}
\end{figure}

\begin{figure}[ht]
\begin{center}
\includegraphics[width=9cm]{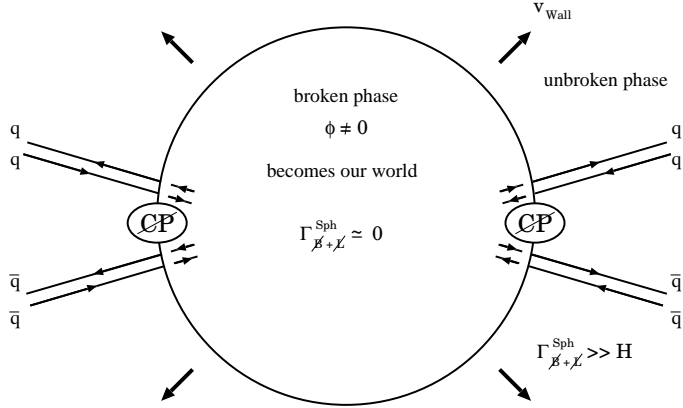} \\ 
\end{center}
\vspace*{-.5cm}
\caption[]{ 
Sketch of nonlocal electroweak baryogenesis.
\protect\label{fig:p44}}
\end{figure}

 As outlined above this scenario works only in extensions of the SM. 
The required departure from
 thermal equilibrium\footnote{The  departure
from thermal equilibrium could have been caused
also by TeV scale topological defects that can arise
in SM extensions \cite{xxTev}.} 
is provided by the expansion of the Higgs bubbles,
the true vacuum. When the bubble walls pass through a  point in space,
the classical Higgs fields change rapidly in the vicinity of such
a point, see Fig. \ref{fig:p46b},
 as do the other fields that couple to those fields.
As far as different mechanisms  are concerned, the following distinction
is made in the literature:\\
$\bullet$ {\it Nonlocal Baryogenesis} \cite{Cohen:1990it},
also called ``charge transport mechanism'',  refers to the case 
where particles and antiparticles  have
CP non-conserving interactions  with a  bubble wall.
This causes an asymmetry
in a quantum number other than B number which is carried 
by (anti)particle currents
into the unbroken phase. There  this asymmetry
is converted by the (B+L)-violating sphaleron processes  into an
asymmetry in baryon number. Some instant later the wall sweeps over
the  region where $\Delta B\neq 0$, filling space with
 Higgs fields that obey  (\ref{eq:vcrit1}). Thus the B-violating
back-reactions are blocked and the asymmetry in baryon mumber persists.
The mechanism is illustrated in Fig. \ref{fig:p44}. \\
$\bullet$ {\it Local Baryogenesis} 
\cite{McLerran:1990zh,Turok:1990zg} refers to case where the
both the  CP-violating and B-violating processes 
occur at or near the bubble walls.\\
In general, one may expect that  both mechanisms were at work  and
$\Delta B \neq 0$  was produced by their joint effort. Which one of 
the mechanisms is more effective depends on the shape and velocity of
the bubbles; i.e., on the underlying model of particle  physics and
its parameters. 
\par
In the following we discuss  only the nonlocal baryogenesis
mechanism.  First, the case of Higgs sector CP violation is treated in
some detail. For definiteness,  we choose a 2-Higgs doublet extension
of the SM 
with CP violation as decribed above. Then (\ref{cphibu}) applies.
Because $|m_\psi(z)|$ becomes, at $T=0$,  the mass of the fermion
 $\psi$, top quarks and, as far as leptons are concerned, $\tau$
leptons have the strongest interactions with the wall. 

\par
\begin{figure}[ht]
\begin{center}
\includegraphics[width=9cm]{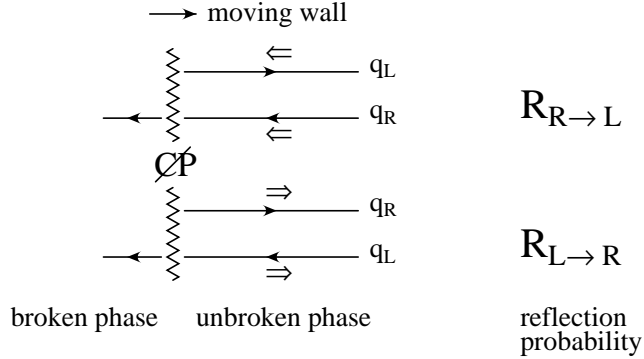} \\ 
\end{center}
\vspace*{-.5cm}
\caption[]{ 
Reflection of left- and right-handed quarks  at a radially expanding
Higgs bubble. The transmission of (anti)quarks from the broken into
the symmetric phase is not depicted.
\protect\label{fig:p46a}}
\end{figure}

We consider for simplicity only the
so-called thin wall regime which applies if the mean free path
of a fermion, $l_\psi$, is larger than the thickness $l_{wall}$. Then 
the quarks and leptons can be treated as free particles,  
interacting only in a small region with a non-trivial Higgs background
field, see Fig. \ref{fig:p46b}.
Multiple scattering within the wall may be neglected. 
The expansion of the wall is supposed to be spherically symmetric
and the 1-dimensional description as given in section 5.3.1 applies.
Fig. \ref{fig:p46a} shows left-handed and
right-handed
quarks\footnote{In this subsection  the symbols $q_L$, ${\bar q_L}$, etc.
 do {\it not}  denote fields but particle states.} 
$q_L$ and $q_R$ incident from  the
unbroken phase, which hit the moving wall and are reflected by the
 Higgs bubble into
 right-handed and left-handed quarks, respectively. 

In the frame where the wall is
at rest, the fermion interactions with the bubble wall are described
by the Dirac equation following from  (\ref{cphibu}):
\begin{equation}
(i\gamma^\mu\partial_\mu -m(z)P_R - m^*(z)P_L)\psi(z,t) \: = \: 0 \: ,
\label{Direqa}
\end{equation}
where $P_{R,L}=(1\pm\gamma_5)/2$ and 
$\psi$ is a  c-number Dirac spinor. Solving this equation 
with the appropriate
boundary conditions yields the (anti)quark wave functions of either
chirality \cite{Nelson:1991ab,Funakubo:en}.

\begin{figure}[ht]
\begin{center}
\includegraphics[width=9cm]{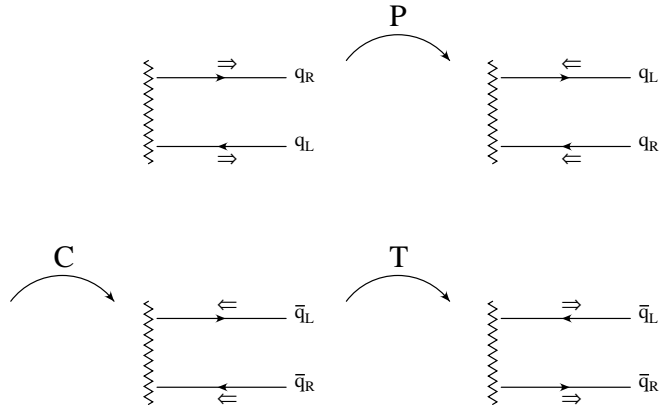} \\ 
\end{center}
\vspace*{-.5cm}
\caption[]{ 
The reflection $q_L\to q_R$ and
the P-, CP-, and CPT-transformed process.
\protect\label{fig:p47}}
\end{figure}

\par
Instead of performing this calculation let's make  a few general
considerations.
Let's have a look at the scattering
process   depicted in Fig. \ref{fig:p47},
where, in 
the symmmetric phase $z>z_0$, a left-handed quark $q_L$ (having
momentum  $k_z<0$)   is reflected at the wall
into  a right-handed $q_R$. Notice that
conservation of electric charge guarantees that a quark is reflected
into a quark and not an antiquark.  Angular momentum
conservation tells us that $q_L$ is reflected as $q_R$ and vice
versa.
Also shown  are the situations after a parity
transformation  (followed by  a
rotation around the wall axis in the paper plane orthogonal to 
the $z$ axis  by an angle $\pi$), and subsequent charge conjugation C,
and time reversal (T)  transformations. The analogous figure  can  be drawn
for antiquark reflection. These figures immediately tell us that if
CP were conserved then 
\begin{equation}
{\cal R}_{L\to R}\: =\: {\cal R}_{{\bar R}\to{\bar  L}} \:,
\qquad {\cal R}_{R\to L}\: =\: {\cal R}_{{\bar L}\to{\bar  R}} 
\qquad 
\label{cprcon}
\end{equation}
would hold. (The subscripts ${\bar R}$, $\bar L$
denote right-handed and left-handed antiquarks, respectively.)
CPT invariance, which is respected by the particle physics
models  we consider, implies 
\begin{equation}
{\cal R}_{L\to R}\: =\: {\cal R}_{{\bar L}\to{\bar  R}} \:,
\qquad {\cal R}_{R\to L}\: =\: {\cal R}_{{\bar R}\to{\bar  L}} \: .
\qquad 
\label{cptcon}
\end{equation}
\par
The charge transport mechanism \cite{Nelson:1991ab}
works  as follows. At some initial time we have equal numbers  of
quarks and antiquarks in the unbroken phase, in 
particular equal numbers of $q_L$ and ${\bar
q}_R$ and $q_R$ and ${\bar
q}_L$, respectively, which hit the expanding bubble
wall.  Reflection 
converts $q_L\to q_R$, ${\bar q}_R  \to {\bar q}_L$,
$q_R\to q_L$, and ${\bar q}_L  \to {\bar q}_R$ and the particles
move back to the region where the Higgs fields are zero. 
Because the
interaction with the bubble wall is  assumed to be
CP-violating, the relations (\ref{cprcon}) for the reflection
probabilities no longer hold.
Actually, for  the CP asymmetry
\begin{equation}
\Delta{\cal R}_{CP}\: \equiv \: {\cal R}_{{\bar L}\to{\bar  R}} - {\cal
  R}_{R\to L}
 \:  =\:
 {\cal R}_{{L}\to{R}} - {\cal R}_{{\bar R}\to{\bar  L}} 
\label{cpadif}
\end{equation}
to be non-zero it is essential that $m_q(z)$ has a $z$ dependent phase.
The reflection coefficients are built up by the coherent superposition
of the amplitudes for (anti)quarks to reflect at some point $z$ in the
bubble. When the phases vary with $z$  the reflection
probabilities  ${\cal R}_{{\bar L}\to{\bar  R}}$ and ${\cal
  R}_{R\to L}$ differ from each other. If the phase of $m_q(z)$ were
constant these probabilities would be equal. (Keep in mind that we work
at the level of 1-particle quantum mechanics.) An explicit computation
yields \cite{Joyce:1994zn}
\begin{equation}
\Delta{\cal R}_{CP}(k_z) \:\propto \:\int_{-\infty}^{+\infty}
 dz \, \cos(2k_z z) {\rm Im}[m_q(z)M_q^*] \:, 
\label{yopt}
\end{equation}
where $M_q$ = $m_q(z=-\infty)$ is the mass of the quark in the broken
phase, and $arg(M_q)=\xi_1$ -- see eq. (\ref{ascpp}). This equation
corroborates the above statement; if $m_q(z)$ had a constant phase,
the asymmetry would be zero. 
Notice  that at this stage
the net baryon number is still zero. This is  because 
the difference $J_q^L$ of the fluxes of $\bar q_R$ and $q_L$, injected
from the wall back into the symmetric phase, is
equal\footnote{Interactions with the
other  plasma particles are neglected.}  to  $J_q^R$ which we
define as  the 
difference  of the fluxes of  $q_R$ and  ${\bar q}_L$,
as should be clear from (\ref{cptcon}). 
However,  the (B+L)-violating weak sphaleron
interactions, which are unsuppressed in the symmetric phase away from
the wall, act  only on the (massless) left-handed quarks and right-handed
antiquarks. For instance, the reaction (\ref{eq:blreact}) decreases the baryon
number by 3 units, while the corresponding reaction with right-handed
antiquarks in the initial state increases B by the same amount.
Thus if the functional form of the CP-violating 
part ${\rm Im}[m_q(z)M_q^*]$ of the 
background Higgs field is such that   $J_q^L>0$ 
then, after the anomalous weak interactions took place, there are more
left-handed  quarks than right-handed antiquarks. 
The   fluxes of the reflected 
${\bar q}_L$ and   $q_R$ 
are  not affected by the anomalous weak
sphaleron interactions. Adding it all up we see that some place away
from the wall a net baryon number $\Delta B>0$ 
is produced. Some instant later
the expanding bubble sweeps over that region and  the associated
non-zero Higgs fields strongly suppress the  (B+L)-violating back
reactions that would
wash out $\Delta B$. Thus the non-zero B number 
produced  before is  frozen in. 
\par
We must also take  into account 
that (anti)particles in the broken phase can be
transmitted  into the symmetric phase and contribute to
the (anti)particle fluxes discussed above. Using CPT invariance and 
unitarity, we find that the probabilities for
transmission and the above reflection probabilities are 
related: 
\begin{eqnarray}
{\mathcal T}_{L\to L}\: = \: 1 -{\cal R}_{R\to L}\: = \: 1 -
{\cal R}_{{\bar R}\to{\bar L}}\: = \: {\mathcal T}_{{\bar L}\to {\bar
    L}} \: , \\
{\mathcal T}_{R\to R}\: = \:  1 -{\cal R}_{L\to R}\: = \:  1 -
{\cal R}_{{\bar L}\to{\bar R}}\: = \: {\mathcal T}_{{\bar R}\to {\bar
    R}} \: .  
\label{trapr}
\end{eqnarray}
We can now write down a formula for the current  $J_q^L$, which
we  define as  
the difference of the fluxes of $\bar q_R$ and $q_L$, injected 
from the wall into the symmetric phase. The contribution from the
reflected particles involves the term $\Delta{\cal R}_{CP}f_s$ where 
$f_s$ is the free-particle Fermi-Dirac phase-space distribution of the
(anti)quarks in the region $z>z_0$ that
move to the left, i.e., towards the wall. 
The contribution from the (anti)quarks which
have returned from the broken phase involves
$({\cal T}_{{\bar R}\to {\bar R}}-{\cal T}_{L\to L})f_b=
-\Delta{\cal R}_{CP}f_b$, where $f_b$ is the 
phase-space distribution of the
transmitted (anti)quarks that
move to the right. The reference frame is the wall frame.
Notice that 
$f_s$ and $f_b$ differ because the wall moves with a velocity
$v_{wall}\neq 0$ -- in our convention from left to right.
The current  $J_q^L$ is given by
\begin{equation}
J_q^L\: =\: \int_{k_z<0}\, \frac{d^3k}{(2\pi)^3}\frac{|k_z|}{E}
(f_s\, -\, f_b)\Delta{\cal R}_{CP} \, ,
\label{jql}
\end{equation}
where $|k_z|/E$ is the group velocity.
The current is non-zero because two of the three Sakharov conditions,
CP violation and departure from thermal equilibrium,  are met. The
current would vanish  if the
wall were at rest in the plasma frame -- which leads  to thermal
equilibrium --,  because then $(f_s-f_b)=0$.
\par 
The current $J^L=\sum_\psi J_{\psi}^L$  is the source for baryogenesis
some distance away from the wall as sketched above. 
We skip the analysis of diffusion and of the conditions under which
local thermal equilibrium is maintained in front of
the bubble wall \cite{Nelson:1991ab,Joyce:1994zn,Cline:1995dg}. 
This determines the densities of the left-handed
quarks and right-handed antiquarks and their associated chemical 
potentials. The rate of
baryon production per unit volume is determined by the equation
\cite{Joyce:1994zn}
\begin{equation}
\frac{dn_B}{dt}\:=\: -n_F \frac{\hat\Gamma_{sph}}{2T}\sum_{generations}
(3\hat\mu_{U_L}+3\hat\mu_{D_L} +
\hat\mu_{\ell_L}+ \hat\mu_{\nu_L}) \: , 
\label{rateq}
\end{equation}
where $n_F=3$, $U=u,c,t$, $D=d,s,b$, 
${\hat\Gamma_{sph}}$ is the sphaleron rate per unit volume,
which in the unbroken phase is given by eq. (\ref{eq:sprabo}).
Here the $\hat\mu_i=\mu_i -{\bar\mu_i}=2\mu_i$ denote 
the difference between the respective particle
and antiparticle chemical potentials.
For a non-interacting gas of massless fermions $i$, 
the relation
between $\hat\mu_i$  and the asymmetry in the corresponding
particle and antiparticle
number densities is
 $n_i-{\bar n}_i \simeq g{\hat\mu_i}T^2/12$, where $g=1$ for a left-handed
lepton and $g=3$ for a left-handed quark
because of three colors. 
In the symmetric phase (\ref{rateq}) then reads
\begin{equation}
\frac{dn_B}{dt}\:=\: -6 n_F \frac{\hat\Gamma_{sph}}{T^3}
(3B_L +L_L) \: , 
\label{rateq1}
\end{equation}
where $B_L$ and $L_L$ denote the total left-handed baryon and lepton
number densities, respectively. The factor of 3 comes from the
definition of baryon number, which assigns baryon number 1/3
to a quark. 
This equation tells us what we already concluded  qualitatively
above:  baryon rather than antibaryon production requires 
a  negative left-handed  fermion number density, i.e.,  a positive
flux  $J_q^L$. 
The total flux $\sum_\psi J_\psi^L$ determines 
the left-handed fermion
number density. Then eq. (\ref{rateq1}) yields $n_B$ and, using
$s=2\pi^2g_{*s}T^3/45$ with $g_{*s}\simeq 110$ 
(see section 2.2), a prediction for the
baryon-to-entropy ratio is obtained.  
\par
So much to the main  aspects
of the mechanism. There are, however, a number of issues that
complicate this scenario. Decoherence effects during
reflection should  be studied. The propagation of fermions is
affected by the ambient high temperature plasma leading to
modifications of their vacuum dispersion relations. 
The shape and velocity of the wall is a critical issue. We refer
to the quoted literature for a discussion of these and other points.
\par
Because Higgs sector CP violation as discussed above is
strongest for top quarks,  one might expect that these quarks
make the dominant contribution to the right hand side of
(\ref{rateq1}). However, several effects tend to decrease their
contribution relative to those of $\tau$ leptons. As top quarks
interact much more strongly than $\tau$ leptons they have a shorter
mean free path. This means that for typical wall thicknesses the 
thin-wall approximation does not hold for $t$ quarks. Further the
injected left-handed
top current $J_t^L$ is affected by QCD sphaleron fields which induce  
processes -- unsuppressed at high $T$ -- where the chiralities of the
quarks are flipped \cite{Mohapatra:1991bz,Giudice:1993bb}. This damps 
the $t$ quark  contribution to $B_L$. 
Refs. \cite{Joyce:1994zn,Cline:1995dg} come to the conclusion that in
this type of  particle physics models 
the contribution of $\tau$ leptons to the left-handed fermion number
density is the most important one. Ref. \cite{Cline:1995dg} finds that  this
induces  a baryon-to-entropy ratio of about
\begin{equation}
\frac{n_B}{s}\: \simeq \: 10^{-12} \frac{\Delta\theta}{v_{wall}} \: ,
\end{equation}
where $v_{wall}$ is the velocity of the wall and
  ${\Delta\theta}\simeq \theta(z=-\infty)-\theta(z=+\infty)$.
Barring the possibility of spontaneous CP violation at non-zero
temperatures in the 2-Higgs doublet models, 
$\Delta\theta$ should be roughly of the order of the
CP-violating phase $\xi$ in the 2-doublet potential (\ref{eq:v2hdm}).
Using that primordial nucleosynthesis allows $n_B/s\simeq \eta/7
\simeq (2 - 8)\times 10^{-11}$  (cf. (\ref{eq:etab})) 
one gets  the parameter  constraint
$\Delta\theta/v_{wall}\sim 40$. Even  large CP violation, 
$\Delta\theta$ of order 1, would require  small wall
velocities, which might not be supported by
investigations of the dynamics of the phase transition. 
Nevertheless, the 2-Higgs doublet models  predict roughly
the correct order of magnitude. In view of the
complexity of this baryogenesis scenario, there are possibly 
additional, hitherto  unnoticed
effects that may influence $n_B/s$. 
For a treatment of the case when the bubble walls are thick, in 
the sense that 
fermions interact with the plasma many times as the wall sweeps
through, see \cite{Joyce:1994zt}.
\par
Only a few words on electroweak baryogenesis in the minimal
supersymmetric standard model,  see e.g.
 \cite{Huet:1995sh,Riotto:1998zb,Cline:2000nw,Carena:2000id}. 
The essentials of the scenario are
analogous to
the 2HDM case, with CP-violating sources as described in section
5.3.2,
the main source for baryogenesis being the phase $\varphi_\mu$ of
the complex Higgsino mixing parameter $\mu_c$.  A number 
of authors conclude 
that the dominant baryogenesis source comes from the
Higgsino sector, which produces a non-zero flux of left-handed quark
chirality.  
The  results for  $n_B/s$  may be presented in the form 
\begin{equation}
\frac{n_B}{s}\: = \: 4 \times 10^{-11} a \sin\varphi_\mu  \: .
\end{equation}
There is  a considerable
 spread in the predicted values of  $a$, respectively in the resulting
estimates of the necessary magnitude of $ \sin\varphi_\mu$.   While
refs. \cite{Riotto:1998zb,Carena:2000id}  find 
that a small  CP phase $\varphi_\mu\gtrsim 0.04$ would suffice to
obtain the correct order of magnitude of  $n_B/s$ (which 
requires, however,
small wall velocities), ref. \cite{Cline:2000nw} concludes  
that $\sin\varphi_\mu$ must be of order 1. 
Large values of $\varphi_\mu$ , however, tend to be in conflict with the
constraints from the experimental upper bounds on the electric dipole
moments of the electron and neutron, see section 5.3.2. 
Electroweak baryogenesis in a next-to-minimal SUSY model was
investigated in \cite{Huber:2000mg}.

\subsection{Role of the KM Phase}
We haven't yet discussed which role is played
in baryogenesis scenarios by the SM source of CP violation, the
KM phase $\delta_{KM}$. 
This question was put  out of
the limelight  after it had become clear 
that the SM alone cannot explain the BAU,  for reasons outlined above. 
Therefore,  SM extensions must be invoked, and such extensions usually
entail in a natural way new sources of CP violation which 
can be quite effective, as far as their role in baryogenesis
scenarios is concerned, as we have seen -- see  also the next section. 
Nevertheless, this is a very relevant issue.
\par
Recall the following
  well-known features of KM CP violation.
All CP-violating effects, which are generated by the KM
phase in the charged weak quark current couplings to W bosons, are
proportional to the invariant \cite{Jarl,Bra}:
\begin{equation}
J_{CP} \: = \: \prod_{i>j \atop u,c,t} (m^2_i - m^2_j) 
\prod_{i>j \atop d,s,b} (m^2_i - m^2_j) 
\quad{\rm Im} \, Q \: ,
\label{JJ}
\end{equation}
where i,j = 1,2,3 are generation indices, $m_u$, etc.   denote the 
respective quark masses, and ${\rm Im}\, Q$ is the imaginary part
of a  product of 4 CKM matrix elements, which is invariant under 
phase changes of the quark fields. There are a number of 
equivalent choices for ${\rm Im}\, Q$. A standard choice is
\begin{equation}
{\rm Im}\, Q = {\rm Im}(V_{ud}V_{cb}V^*_{ub}V^*_{cd}) \: . 
\label{ImQ}
\end{equation}
Inserting the moduli of the measured CKM matrix elements
yields $|{\rm Im}\, Q|$ smaller than $2\times 10^{-5}$, even 
if KM CP violation is
maximal; i.e., $\delta_{KM}=\pi/2$ in the KM parameterization of
the CKM matrix. We may write 
${\rm Im}\, Q\simeq 2\times  10^{-5}\sin\delta_{KM}$. 
As far as the SM at temperatures $T\neq 0$ is concerned, the CP symmetry
can be broken  only in regions of space where the gauge symmetry is
also broken, or at the boundaries of such regions, because 
$J_{CP}\neq 0$ requires non-degenerate quark masses. 
Imagine the EW transition would be first  order due to a 2-Higgs
doublet extension of the SM with {\it no} CP violation in the Higgs
sector.
The question is then: is the KM source of CP violation strong enough
to create a sufficiently large asymmetry $\Delta{\cal R}_{CP}$
in the  probabilities for reflection of (anti)quarks at the expanding
wall as discussed above? It is clear that  $\Delta{\cal R}_{CP}$
must be proportional to a dimensionless quantity of
the form $J_{CP}/D$, where $D$ has mass dimension 12. Reflection of 
quarks and antiquarks at a bubble wall is not CKM-suppressed; hence 
$D$ does not contain small CKM matrix elements. If one recalls that
in the symmetric  phase the quark masses and thus $J_{CP}$ vanish, it
seems reasonable to treat the quark masses (perhaps not the top quark
mass) as a perturbation. In the massless limit the mass scale of the
theory at the EW transition is then given by the critical temperature
$T_c\sim$ 100 GeV. Thus one  gets for the dimensionless measure
of CP violation:
\begin{equation}
d_{CP}\: \equiv\: \frac{J_{CP}}{T_c^{12}} \:
 \sim \: 10^{-19} \:  
\label{dimea}
\end{equation}
as an estimate of $\Delta{\cal R}_{CP}$. Clearly this number is
orders of magnitude too small to account for the observed $n_B/s$.
CP violation \`a la KM is therefore classified, by consensus of opinion,
as being irrelevant for baryogenesis.
It was argued, however, that there  may
exist significant enhancement effects \cite{Farrar:hn},
and there has been a considerable debate over this issue
\cite{Farrar:hn,Gavela:ds,Huet:1994jb}.

\section{Out-of-Equilibrium Decay of Super-Heavy Particle(s)}
Historically the first type of
baryogenesis scenario  which was developed in
detail is the so-called out-of-equilibrium decay of a super-heavy
particle 
\cite{Yoshimura:1978ex,Dimopoulos:1978kv,Toussaint:1978br,Weinberg:1979bt}.
The basic idea is that at a very early stage of the
expanding universe, a super-heavy particle species $X$ existed, the
reaction rate of which became  smaller than the expansion rate $H$ of the
universe at temperatures $T\gg T_{EW}$. Therefore, 
these particles decoupled from the
thermal bath and became over-abundant. The decays  of $X$ and of the
antiparticles $\bar X$  are supposed to be CP- and B-violating, such that a net
baryon number $\Delta B\neq 0$ is  produced when the $X,\bar X$
have decayed. This scenario has its
natural setting in the framework of grand unified theories. A brief
outline is given in the next subsection.
A viable variant is baryogenesis via leptogenesis through the lepton-number
violating decays of (a) heavy Majorana neutrino(s)
\cite{Fukugita:1986hr}. This will be discussed  in subsection 6.2.

\subsection{GUT Baryogenesis}
The ``out-of-equilibrium decay'' scenario is natural
in the context of grand unified theories. Grand unification aims at
unifying the strong and electroweak interactions at some high energy
scale. It works, for instance, in the context of supersymmetry, where
the effective couplings of the strong, weak, and electromagnetic
interactions
become equal at an energy scale $M_{GUT}\simeq 10^{16}$ GeV
\cite{Giudice}. 
A matter multiplet forming a representation of the GUT gauge group $G$
contains both quarks and leptons.  Gauge bosons mediate transitions
between
the members of this multiplet, and -- for many  gauge groups 
-- some of the gauge bosons 
induce B-violating processes. Also C violation 
and non-standard CP violation occurs naturally. As to the latter: the 
gauge group $G$ must be broken at the GUT scale 
$M_{GUT}\simeq 10^{16}$ GeV to some smaller symmetry group
$G'\supseteq SU(3)_c\times SU(2)_L\times U(1)_Y$. This is accomplished
by scalar Higgs multiplets. As a consequence, GUTs contain in general
super-heavy Higgs bosons with B-violating and CP-violating Yukawa
couplings to quarks and leptons. 
\par
The simplest example of a GUT is based on the gauge group $G=SU(5)$. It is
obsolete  because the model is  in conflict with the stability of the proton. 
Irrespective of this obstruction the minimal version of this  model
is  of no use for implementing the scenario which we discuss below,
because the interactions of minimal $SU(5)$ conserve $B-L$. A popular
gauge group is $SO(10)$ which allows to construct models that 
avoid both obstacles \cite{Giudice}. 
\par
Rather than going into the details of a specific GUT let us
illustrate the baryogenesis mechanism with a well-known toy
 model \cite{Kolb}.
Consider a super-heavy leptoquark gauge boson $X$
which is supposed to have   quark-quark and  quark-lepton
decay channels, $X\to qq,
{\ell} {\bar q}$. In the table the branching ratios of these
decays, of the decays of the antiparticle $\bar X$, and the
baryon numbers $B$ of the final states are tabulated.
\begin{center}\renewcommand{\arraystretch}{1.5}
\begin{tabular}{c|c|c} 
 final state $f$ & branching ratio & $ B$   \\ \hline
$X\to qq$ & $r$   & 2/3 \\ 
$X\to {\ell}{\bar q}$ & $1-r$ & -1/3  \\ 
 ${\bar X} \to {\bar q}{\bar q}$ & ${\bar r}$ & -2/3 \\
 ${\bar X} \to {\bar\ell}{q}$ & ${1-\bar r}$ & 1/3 \\
\end{tabular}
\end{center}

\begin{figure}[ht]
\begin{center}
\includegraphics[width=9cm]{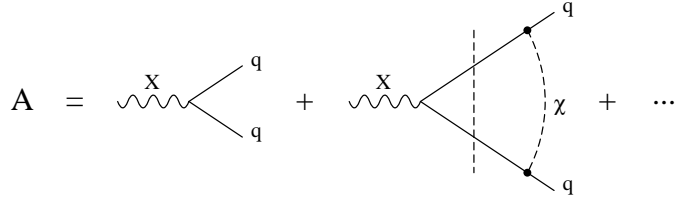} \\ 
\end{center}
\vspace*{-.5cm}
\caption[]{ 
Amplitude for the decay $X\to qq$ to one-loop approximation. The
vertical dashed line indicates the absorptive part of the 1-loop
contribution which enters $\Delta B_X$. 
\protect\label{fig:p33}}
\end{figure}

\noindent
The baryon number produced in the decays of $X$ and $\bar X$ is:
\begin{eqnarray}
B_X \: =\: \frac{2}{3} r -\frac{1}{3}(1-r) \: , \nonumber \\
B_{\bar X} \: =\: -\frac{2}{3} {\bar r} +\frac{1}{3}(1-{\bar r}) \: ,
\end{eqnarray}
and the net baryon number produced is 
\begin{eqnarray}
\Delta B_X & \equiv &  B_X + B_{\bar X} = r - {\bar r} \nonumber \\
& = & \frac{\Gamma(X\to f_1)}{\Gamma_{tot}(X)} - 
\frac{\Gamma({\bar X}\to {\bar f_1})}{\Gamma_{tot}(\bar X)} 
= \frac{\Gamma(X\to f_1) -\Gamma({\bar X}\to {\bar f_1})}{\Gamma_{tot}}
\: ,
\end{eqnarray}
where $f_1=qq$ and we have used that ${\Gamma_{tot}(X)}=
{\Gamma_{tot}(\bar X)}$,  which follows from CPT invariance. 
Obviously if C or CP were conserved then
$\Delta B_X =0$.  Suppose the quarks and  leptons $q,\ell$ 
couple to a spin-zero boson
$\chi$ with  Yukawa couplings  that contain a non-removable
 CP-violating phase. It is natural to assume that
C is already violated  in the tree-level interactions
of  $X$ to fermions. 
The CP-violating interactions affect 
the $X,\bar X$ 
decay amplitudes beyond the tree level, as shown in Fig. \ref{fig:p33}.
The decay amplitude for $X\to qq$ is, up to spinors and a polarization
vector describing the external particles, 
\begin{equation}
A(X\to qq) \:  =\: A_0 + A_{1}= A_0 + B e^{i\delta_{CP}} \: , 
\label{agut1}
\end{equation} 
where the tree amplitude $A_0$ is real. 
$\Delta B_X \neq 0$ requires, in addition  to CP violation, also a
non-zero final-state
interaction phase. Therefore, the masses of the $X$ boson and of the
fermions must be such that the  intermediate fermions 
in the 1-loop contribution to the amplitude  can be on their 
respective mass shells and re-scatter to produce the final state. This
causes
a  complex $B=|B|\exp(i\omega)$.
The decay amplitude for ${\bar X}\to {\bar q}{\bar q}$ is
\begin{equation}
A({\bar X}\to {\bar q}{\bar q}) \: =\:  A_0 + |B|e^{i\omega}
e^{-i\delta_{CP}} \: . 
\label{agut2}
\end{equation} 
Using  (\ref{agut1}), (\ref{agut2}) one obtains that
\begin{equation}
 \Delta B_X \: \propto \:
 \frac{|AB|\sin\omega\sin\delta_{CP}}{\Gamma_{tot}} \: , 
\label{agut3}
\end{equation} 
and the constant of proportionality includes factors from phase 
space integration.\footnote{In charged B meson decays a  CP
asymmetry $A_{CP}=[\Gamma(B^+\to f)-\Gamma(B^-\to \bar f)]/
[\Gamma(B^+\to f)+\Gamma(B^-\to \bar f)]
$ arises in completely analogous fashion. $A_{CP}\neq 0$ requires 
CP violation and final state interactions.}
In addition, the baryon number $\Delta B_{\chi}$
 produced in the decays of the
$\chi,\bar\chi$ bosons must also be computed. Let's assume
 that  $\Delta B_X+\Delta B_{\chi}\neq 0.$

\par
As long as the interactions of these bosons, which include decays, inverse
decays, annihilation,  the B-violating reactions 
$ qq \to \ell{\bar q}$, ${\bar q}{\bar
q} \to {\bar\ell}{q}$, etc. (remember the discussion in section 3.2) are fast
compared to the expansion rate $H$,  the $X,\bar X, \chi, \bar \chi$ have 
thermal distributions and the average baryon number  of the plasma 
remains zero.
Therefore the interactions of these bosons must be weak enough that
they can fall out of equilibrium.
 This is a delicate issue,
because these particles carry gauge charges and
can couple quite strongly to the plasma of the early universe.
\par
Let's outline the scenario for the  $X,\bar X$. (It applies also to the scalar
particles.)
At temperatures $T\gg
m_X$, where $m_X$ is the mass of the $X$ boson, 
the $X,{\bar X}$ particles have relativistic velocities and are
assumed
to be in thermal equilibrium. Then $n_X=n_{\bar X}\sim n_{\gamma}$
holds
for their number densities. At lower temperatures, the $X,{\bar X}$ bosons
become
non-relativistic and, as long as they remain in thermal equilibrium,
their densities get  Boltzmann-suppressed with decreasing temperature,
$n_{\bar X}=n_X\sim (m_XT)^{3/2}\exp(-m_X/T)$. 
Because 
$\Gamma_{annih}\propto n_X$, the total 
rate $\Gamma_X\sim\alpha m_X $ of $X$ and $\bar X$ decay 
is the relevant number to 
compare with $H$.  If  $\Gamma_X<H$, an excess of  
$X,\bar X$  with respect to their equilibrium numbers will develop.
The $X,\bar X$ drift
along in the expanding universe for a little while and decay. Notice
that
the inverse decays, $f\to X$, ${\bar f}\to {\bar X}$, by which bosons
are
created again by quark-quark annihilation, etc. are blocked, because
the fraction of these fermions with sufficient energy to produce a
super-heavy boson is Boltzmann suppressed for $T<m_X$. At the time
of their decay, $t\sim \Gamma_X^{-1}$, there is quite an
over-abundance: $n_X=n_{\bar X}\sim n_{\gamma}(T_{decay})$. Using that
the entropy density $s\sim g_*n_{\gamma}$ (see sect. 2.2) one gets for
the produced baryon asymmetry:
\begin{equation}
\frac{n_B}{s} \: \sim \: \frac{\Delta B \,  n_{\gamma}}{ g_*n_{\gamma}}
\: \sim \: \frac{\Delta B}{g_*}  \: ,
\label{gutnb}
\end{equation}
where $\Delta B \simeq \Delta B_X$ is the baryon number produced per
boson decay.  For a (GUT) extension of 
the SM one may expect that  $g_*$ is somewhere between $10^2$ and
$10^3$. Thus only a tiny CP asymmetry  $\Delta B\sim 10^{-8}-10^{-7}$
is required to obtain $n_B/s\sim 10^{-10}$. Of course, these crude
estimates must be made quantitative by computing the relevant reaction
rates using a specific particle physics model, and tracking
the time evolution of the particle densities by solving the
Boltzmann equations. A detailed exposition is given in  \cite{Kolb}.
\par 
The above condition for the decoupling of the $X$ particles from the
thermal bath, $\Gamma_X\sim\alpha m_X <H$, translates into a condition
on the mass of the spin 1 gauge boson: $m_X > \alpha g_*^{-1/2} m_{Pl}\sim
10^{16}$ GeV, where $\alpha=g_{gauge}^2/(4\pi)\sim 10^{-2}$. For
super-heavy scalar bosons with B-violating decays  a  mass bound
obtains which is lower (cf., e.g., \cite{Riotto:1999yt}). 
\par 
There are several pitfalls that constrain this type of baryogenesis
mechanism. First, remember
that a scenario that tries to explain the BAU by a mechanism that
operates above the temperature $T_{EW}\sim 100$ GeV must 
involve interactions that violate $B-L$. 
In the context of grand unified theories, models based  on the gauge
group $SO(10)$ lead to $(B-L)$ non-conservation. These models have
several attractive features, in particular  with respect 
to the scenario discussed 
in the next subsection.
\par
GUT baryogenesis
may be in conflict with inflation. This is a serious problem. 
 An essential  assumption in the
above scenario was that at very high temperatures $T$ above $m_X$  the
$X,\bar X$ particles were in thermal equilibrium and were 
as abundant as photons. If these  particles  are the super-heavy gauge or
Higgs bosons  of a GUT this assumption may be wrong. It might be
that  the temperature  of the
quasi-adiabatically
expanding plasma of particles in the early universe
  was always smaller than $M_{GUT}$.
There are  a number of  reasons to believe that the energy of the
very early universe was dominated by vacuum energy,  which led to
exponential expansion of the cosmos. This is the basic
assumption of the inflationary model(s). These models  
solve a number of fundamental cosmological problems, including the
monopole
problem. A number of GUTs predict super-heavy, stable magnetic
monopoles. Their contribution to the energy density of the universe
would over-close the cosmos -- but that is not observed. Inflation
would sweep away these monopoles, along with other particles, leaving
an empty space. At the end of inflation 
 the vacuum energy is converted through quantum fluctuations into
pairs of relativistic particles and antiparticles which then
thermalize. This process is called reheating and it can be
characterized
by an energy scale called the reheat temperature $T_r$. If  the
reheating process is fast, i.e., if  $T_r$ is above $M_{GUT}$, the
monopoles are re-created. On the other hand if
 reheating occurred slowly  such that
$T_r$ is well below  $M_{GUT}$ then the re-production of $X\bar X$
super-heavy gauge and Higgs bosons -- which were to  
initiate  baryogenesis as 
described above -- appears to be inhibited,  or should at least be
suppressed. See \cite{Riotto:1999yt} 
for an overview on  ways to circumvent this and associated
problems. 

\subsection{ Baryogenesis through Leptogenesis}
This mechanism is a special case of the ``out-of-equilibrium decay'' scenario.
It assumes the existence of a heavy Majorana neutrino species in the
early universe above $T_{EW}$ with a particle mass, typically, of the order
of $M\sim 10^{12}$ GeV -- or, in fact,  the more realistic
case  of three  heavy Majorana neutrino species with
non-degenerate masses. These particles interact only weakly with the other
particle species in the early universe and fall out of equilibrium at some
temperature $T\sim M \gg T_{EW}$. It is essential that 
some of the interactions of the underlying particle 
physics model do not conserve 
$B-L$. The  heavy Majorana neutrinos decay, for instance
into ordinary leptons and Higgs bosons  which are the most important channels,
thereby generating a non-zero lepton number. 
Lepton-number violating scattering 
processes must not wash out this asymmetry. Then the $(B-L)$-conserving
SM sphaleron reactions, which occur rapidly enough above  $T_{EW}$,
convert this lepton asymmetry into a baryon asymmetry.
\par
The scenario was suggested in \cite{Fukugita:1986hr}, and it has been
subsequently  developed further -- see 
\cite{Luty:un,Flanz:1994yx,Plumacher:1996kc,Covi:1996wh,Buchmuller:1997yu,Pilaftsis:1998pd}
and the reviews \cite{Buchmuller:2000as}. 
The attractiveness of this
scenario stems from the observed atmospheric and solar neutrino deficits
which point to oscillations of the light neutrinos. It is well-known
that these data  can be explained by small differences in the  masses of the
electron, muon, and tau neutrinos. The value of $\Delta m^2_{23}=
m^2_3- m^2_2$ extracted from the data indicates  that the mass of the
heaviest of the three light neutrinos is of the order of $10^{-2}$ eV. 
Such small masses
can be explained in a satisfactory way by the so-called seesaw mechanism
\cite{seesaw}. This mechanism requires (i) the neutrinos to be Majorana
fermions and (ii) three very heavy right-handed neutrinos which are
singlets with respect to $SU(2)_L\times U(1)_Y$ -- see Appendix B. 
\par
Within the framework of GUTs, popular models are based on the gauge group 
$SO(10)$  which contain in their  particle spectra ultra-heavy right-handed
Majorana neutrinos with lepton-number violating decays. We consider here
only a minimal, non-GUT model. Take the electroweak standard model and add
three right-handed $SU(2)_L\times U(1)_Y$ singlet fields $\nu_{\alpha R}$
$(\alpha=1,2,3)$ with a Majorana mass term for these fields involving 
mass parameters much larger than $v=246$ GeV. The general Yukawa interaction
for the charged leptons and neutrinos is then given by eq. (\ref{eq:3gen})
of appendix B with $\Phi\equiv \Phi_1=\Phi_2$, where $\Phi=(\phi^+,\phi^0)^T$
is the SM $SU(2)_L$ doublet field. As described in appendix B we have 
in the mass basis three very light, practically left-handed Majorana 
neutrinos, which we identify with the neutrinos we know, and three very heavy,
right-handed Majorana neutrinos $N_i$. Let's  switch back to the 
early universe  when the  $N_i$ were still around. The 
interaction (\ref{eq:3gen}) implies that the $N_i$ have lepton-number
violating decays at tree-level,  $N_i\to \ell \phi$  and
$N_i\to {\bar\ell} \phi^*$, where $\ell, \phi$ denotes
 either a  negatively charged
lepton and a  $\phi^+$ (which later ends up as the longitudinal component
of the $W^+$ boson) or a light neutrino and a $\phi^0$. C and CP
violation cause a difference in these two rates -- see below. 
As long as the $N_i$ are in thermal equilibrium 
CPT invariance and the unitarity of the S matrix ( cf. section 3.2)
guarantee that the average lepton number remains zero. (The $N_i$ are to
be described as on-shell resonances in corresponding $2\leftrightarrow
2$ processes.)
When the $N_i$ have fallen out of equilibrium, there is still the
danger of lepton-number violating wash-out processes, for instance
${|\Delta L|=2}$ reactions mediated by $N_i$ exchange. The requirement
 $\Gamma_{|\Delta
  L|=2}(T) < H(T)$ for temperatures  $T$  smaller
than the leptogenesis temperature, e.g. $T\lesssim 10^{10}$ GeV, 
 implies an upper bound on the masses of the light neutrinos
\cite{Buchmuller:2000as}. 
\par
\begin{figure}[ht]
\begin{center}
\includegraphics[width=9cm]{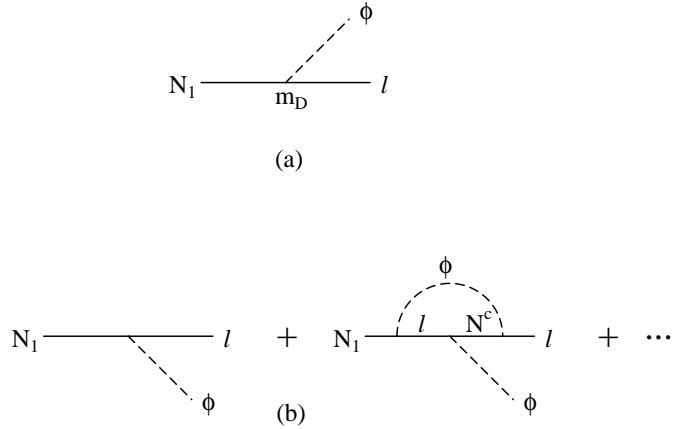} \\ 
\end{center}
\vspace*{-.5cm}
\caption[]{ 
Two-body decay of a super-heavy Majorana neutrino
$N_1\to \ell \phi$: Born amplitude (a) and a 1-loop contribution (b).
Self-energy contributions are  not depicted. 
\protect\label{fig:p56}}
\end{figure}

We assume the $N_i$ to be non-degenerate and put the labels such that
$M_3>M_2>M_1$ holds for the masses. The decay width of $N_i$  at tree
level in its
rest frame, see Fig. \ref{fig:p56}, is easily computed using
(\ref{eq:3gen}):
\begin{eqnarray}
\Gamma_i & \equiv & \Gamma(N_i\to \ell \phi) + 
\Gamma(N_i\to {\bar\ell} \phi^*) \nonumber \\
& & = \frac{(M_D^\dagger M_D)_{ii}}{4\pi v^2} M_i \: , 
\label{gnid}
\end{eqnarray}
where $M_D$ is the Dirac mass matrix (\ref{eq:D3gen}). 
For leptogenesis to work   the decays of the $N_i$ must be slow
as compared to $H$. The condition $\Gamma_i < H(T=M_i)$ is fulfilled
only if the masses of the light neutrinos are small, roughly 
$m_{\nu_i} < 10^{-3}$ eV \cite{Fischler:1990gn}, which is compatible
with observations. There is then an excess of the heavy
neutrinos
with respect to their rapidly decreasing equilibrium distributions 
$n_{eq} \sim \exp(-M/T)$.

\par
 Eventually the $N_i$  decay and  lepton number is 
produced. It is due to the CP-asymmetry in the decay rates which is
 generated by the interference of the tree amplitude with the 
absorptive part of the 1-loop amplitude depicted in  Fig. \ref{fig:p56},
analogous to eq. (\ref{agut3}). If $M_1\ll M_2,M_3$  one obtains
for the decay of $N_1$:

\begin{eqnarray}
\epsilon_1 & \equiv & \frac{\Gamma(N_1\to \ell \phi) - 
\Gamma(N_1\to {\bar\ell} \phi^*)}{\Gamma(N_1\to \ell \phi) +
\Gamma(N_1\to {\bar\ell} \phi^*)}  \nonumber \\
& & \simeq  - \frac{3}{4\pi v^2} 
\frac{1}{(M_D^\dagger M_D)_{11}}\sum_{j=2,3} {\rm Im}
[(M_D^\dagger M_D)^2_{1j}] \frac{M_1}{M_j}\: .
\label{gnid1}
\end{eqnarray}

The asymmetries $\epsilon_i$ are determined by the 
moduli and the CP-violating phases
of the elements of the matrix $M_D^\dagger M_D$.
The moduli are related to the light neutrino masses,  
while the  CP-violating phases  are in general
unrelated to the CP-violating phases of the mixing matrix in the 
leptonic charged current-interactions involving the light neutrinos --
see appendix B. 
\par
\begin{figure}[ht]
\begin{center}
\includegraphics[width=9cm]{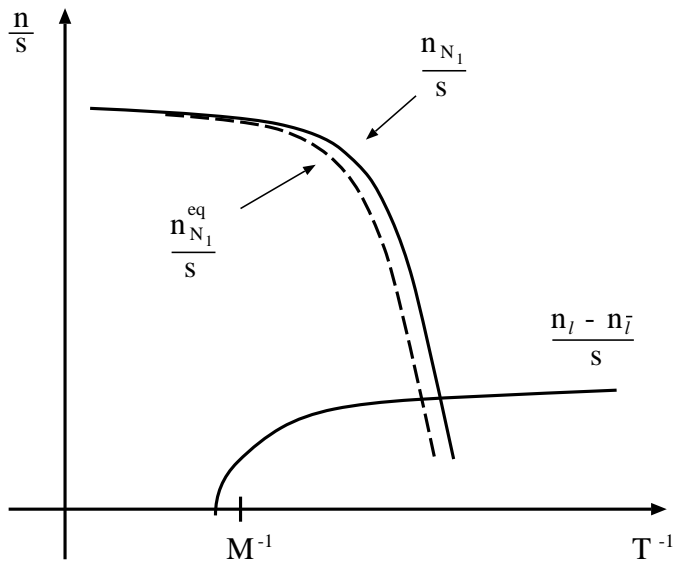} \\ 
\end{center}
\vspace*{-.5cm}
\caption[]{ 
Evolution of the ratio $n_{N_1}/s$ as the universe cools
off. Departure
from thermal equilibrium occurs at
$T\lesssim M_{N_1}$ and a leptonic asymmetry is generated
\cite{Buchmuller:2000as}.
\protect\label{fig:p57}}
\end{figure}

 
The asymmetry (\ref{gnid1}) 
corresponds to an asymmetry in the density of leptons
versus antileptons, $n_L\equiv n_\ell - n_{\bar\ell} \neq 0.$ A crude
estimate of the lepton-number-to-entropy  ratio  $Y_L=n_L/s$  gives
\begin{equation}
{Y_L} \: \sim \: \epsilon_1 \frac{n_N}{s} \:  \sim \:
 \frac{\epsilon_1}{g_*} \: , 
\label{nlse1}
\end{equation}
where $g_* \sim 100$ in the SM. Due to wash-out
processes like those mentioned above 
this ratio is, in fact, smaller than (\ref{nlse1}). In order to
determine the suppression factor $\kappa$, the Boltzmann equations 
for the time evolution of the particle number densities must  be
solved \cite{Luty:un,Plumacher:1996kc,Buchmuller:2000as}.
A typical solution for $n_{N_1}$ is sketched in
Fig. \ref{fig:p57}.
Refs.  \cite{Plumacher:1996kc,Buchmuller:2000as} find 
$\kappa \sim  10^{-1}-10^{-3}$, depending on
the particle physics model.

\par
The asymmetry in lepton number feeds the $(B-L)$-conserving
weak sphaleron reactions, which occur rapidly enough above $T_{EW}$,
and produce an asymmetry in baryon number $Y_B=n_B/s$. There
is a relation between $Y_B$ and the corresponding asymmetries
$Y_{B-L}$ and $Y_L$.   For a given
particle physics model this relation depends 
on the processes which are in thermal equilibrium,  and it is  given
by \cite{Khlebnikov:sr,Harvey:1990qw}:
\begin{equation}
{Y_B} \: = \: C \, Y_{B-L} \:  = \:\frac{C}{C-1}\, Y_L \: . 
\label{hartu}
\end{equation}
The particle reactions which are fast enough  as compared with
$H$ yield relations among the various chemical potentials, and
these relations determine the number $C$. For the minimal model
considered above, one has $C=28/79$ in the high temperature phase
if all but the 
$|\Delta  L|=2$ reactions
are in thermal equilibrium \cite{Khlebnikov:sr}. (In general
$C$ depends on the ratio $v_T/T$, where $v_T$ is the Higgs VEV which
develops in the broken  phase \cite{Khlebnikov:1996vj,Laine:1999wv}.)
Using (\ref{nlse1}),  (\ref{hartu}) the generated baryon-to-entropy
ratio is estimated to be 
\begin{equation}
{Y_B} \: \sim  - Y_L \: =\: - \kappa  \frac{\epsilon_1}{g_*} \: . 
\label{bhartu}
\end{equation}
Using $g_* \sim 100$ and a dilution factor $\kappa \sim 10^{-2}$,
we see that only a very small lepton-asymmetry 
$\epsilon_1\sim 10^{-6}$ is needed. In fact, lepton-number
violation must not  be too strong, in order that the whole 
scenario works. 
\par
Detailed studies of leptogenesis have been made, for
a number of SM extensions and using {\it Ans\"atze} for the neutrino mass
matrices that fit well to the observations concerning
the solar and atmospheric neutrino deficits
\cite{Buchmuller:2000as}. The conclusion is that
${Y_B}\sim 10^{-10}$ is naturally explained by the decay of heavy
Majorana neutrinos, the lightest of which having a mass $M_1 \sim
10^{10}$ GeV, and the required pattern  of the 3 light neutrino
masses  is consistent  with observations.

\section{Summary}
In these lectures I  have outlined two popular theories of baryogenesis, 
which presently seem to be the most plausible ones: electroweak baryogenesis
and out-of-equilibrium decay scenarios, in particular 
baryogenesis via leptogenesis by the decays of ultra-heavy Majorana neutrinos.
A number of other, quite ingenious mechanisms for generating the
BAU  were  conceived. Their
discussion is, however, beyond the scope of these notes and 
the reader is referred to the quoted reviews.  
\par
Electroweak baryogenesis (EWBG) will be testable in the not too distant future.
The clarification of the origin of electroweak symmetry breaking will
be a central physics issue at the Tevatron and at future colliders,
and the outcome will be crucial for the EWBG scenario. 
An important result was already obtained: Theoretical investigations
of the SM electroweak phase transition and the experimental 
lower bound from the LEP experiments on 
 the mass of the SM Higgs boson, $m_H^{SM}>114$ GeV,
led to the conclusion that the standard model of particle
physics fails to explain the BAU. EWBG is still viable in extensions of the
SM, the most popular of which is the minimal supersymmetric extension.
However, the requirement of the electroweak phase transition to be 
strongly first order translates into tight upper bounds on the mass
of the lightest Higgs boson, $m_H < 115$ GeV, and on the mass of the
lighter of the two stop particles, $m_{\tilde t_1} < 170$ GeV, of the MSSM.
Another important ingredient to EWBG is non-standard CP violation.
This motivates the search for T-violating  effects in experiments
with atoms and molecules, and neutrons. Non-SM CP violation can also be traced
in B meson decays or in high-energy reactions including  the production
and decays of top quarks and Higgs bosons, if Higgs  
particles will be discovered.
\par 
GUT-type baryogenesis scenarios cannot be falsified by laboratory
experiments, but they  would, of course, get spectacular
empirical support if 
proton decay would be found, etc. That's what makes leptogenesis by the
decays of ultra-heavy Majorana neutrinos attractive: it has, albeit
indirect,  support
from the observed atmospheric and solar neutrino deficits. Theoretical 
investigations have shown that the scenario is consistent. As far as unknown 
parameters are concerned, the degree of arbitrariness is constrained: in order
to obtain the correct order of magnitude of the BAU the masses of the light
neutrinos must lie in range which is consistent with the interpretation of the 
solar and atmospheric neutrino data. The scenario would get a further push if
the light neutrinos would turn out to be Majorana particles.
Future particle physics experiments and/or astrophysical observations will
bring us closer to understanding what is responsible for the
matter-antimatter asymmetry of the universe.

\subsubsection*{Acknowledgments}
I wish to thank 
M. Beyer and  A. Brandenburg for comments on the manuscript and 
I am indebted to T. Leineweber for the production of  the
postscript figures.
\\ \\ \\
\noindent
{\large \bf Appendix A} \\ \\
Let $q({\bf x},t)$ be the Dirac field operator that 
describes a quark of flavor $q = u,...,t$,
$q^{\dagger}({\bf x},t)$ denotes its Hermitean adjoint, and  ${\bar q} =
q^{\dagger}\gamma^0$. 
The baryon number operator (\ref{eq:bcharge})
is
\begin{equation}
\hat B = \frac{1}{3}\sum_q \int d^3x:q^{\dagger}({\bf x},t)q({\bf x},t): \: ,
\label{eq:bop} 
\end{equation}
and the colons denote normal ordering. Let  C, P   denote 
the unitary and T the anti-unitary operator
which implement  the charge conjugation,  parity, and time reversal
 transformations, 
respectively, in the space of states.
Their action on the  quark fields is, adopting standard phase conventions, 
\begin{eqnarray}
Pq({\bf x},t)P^{-1} & = & \gamma^0q({-\bf x},t) \, , \\
Pq^{\dagger}({\bf x},t)P^{-1} & = &q^{\dagger}({-\bf x},t) \gamma^0 \, , \\
Cq({\bf x},t)C^{-1} & = & i\gamma^2 q^{\dagger}({\bf x},t) \, , \\
Cq^{\dagger}({\bf x},t)C^{-1} & = & i q({\bf x},t)\gamma^2 \, , \\
Tq({\bf x},t)T^{-1} & = & -i\ q({\bf x},-t)\gamma_5\gamma^0\gamma^2 \, , \\
Tq^{\dagger}({\bf x},t)T^{-1} & = & -i \gamma^2
\gamma^0\gamma_5 q^{\dagger}({\bf x},-t) \, , 
\label{eq:c1cpp}
\end{eqnarray}
where $ \gamma^0$, $\gamma^2$, and  $\gamma_5 =
i\gamma^0\gamma^1\gamma^2\gamma^3$
 denote Dirac  matrices. Then
\begin{eqnarray}
P:q^{\dagger}({\bf x},t)q({\bf x},t):P^{-1} \:  = \:
 :q^{\dagger}({-\bf x},t)q({-\bf x},t): \, , \\
C:q^{\dagger}({\bf x},t)q({\bf x},t):C^{-1} \:  = \:
:q({\bf x},t)q^{\dagger}({\bf x},t):
\:  = \: - :q^{\dagger}({\bf x},t)q({\bf x},t): \, , \\
T:q^{\dagger}({\bf x},t)q({\bf x},t):T^{-1} \:  = \:
:q^{\dagger}({\bf x},-t)q({\bf x},-t): \, .
\label{eq:cpdichte}
\end{eqnarray}
With these relations we immediately obtain: 
\begin{eqnarray}
P{\hat B} P^{-1} & = & {\hat B} \: , \\
C{\hat B} C^{-1} & = & -{\hat B} \: .
\label{eq:c2cpp}
\end{eqnarray}
As shown in section 4 the baryon number operator is time-dependent
due to non-perturbative effects. Using translation invariance we have
${\hat B(t)}=e^{iHt}{\hat B(0)}e^{-iHt}$, where $H$ is the
Hamiltonian
of the system. The operator ${\hat B(0)}$ is even 
with respect to  T and odd with respect to  $\Theta \equiv CPT$:
\begin{equation}
\Theta{\hat B(0)} {\Theta}^{-1}  \:  =  \:  -{\hat B(0)} \: .
\end{equation} 
 \\ \\ 
{\large \bf Appendix B} \\ \\
Here we discuss the general structure of $SU(2)_L\times U(1)_Y$
invariant Yukawa interactions in the lepton sector if neutrinos are
Majorana particles. Let's first collect some basic formulae for
Majorana fields. Consider a Dirac field
\begin{equation}
\psi(x) \: =\: \left (\begin{array}{c} \xi(x) \\ \eta(x) \end{array}
  \right ) \, ,
\label{eq:psi}
\end{equation}
where $\xi,\eta$ are 2-component spinor fields. In the chiral representation
of the $\gamma$ matrices, using the
convention where $\gamma_5 =
 {\rm diag}({\rm I}_2,
-{\rm I}_2)$, we have $\xi=\psi_R, \eta=\psi_L$, where $\psi_R,\psi_L$
are the right-handed and left-handed Weyl fields. In the chiral
representation the charge conjugated spinor field $\psi^c$  reads
\begin{equation}
\psi^c \: \equiv\:  i\gamma^2 \psi^{\dagger}
 \: = \: \left (\begin{array}{c} i\sigma_2\eta^{\dagger} \\ 
-i\sigma_2\xi^{\dagger}  \end{array} \right ) \, ,
\label{eq:psic}
\end{equation}
and $\sigma_2$ is the second  Pauli matrix. Let's use the Weyl fields in
4-component form, $\psi_R=(\xi,0)^T, \psi_L=(0,\eta)^T$, and determine,
using (\ref{eq:psic}),  their charge-conjugates:
\begin{eqnarray}
\psi^c_L \: & \equiv & \: (\psi_L)^c \: = \: 
 \left (\begin{array}{c} i\sigma_2\eta^{\dagger} \\ 
0  \end{array} \right ) \, , \\
\psi^c_R \: & \equiv & \: (\psi_R)^c \: = \: 
 \left (\begin{array}{c} 0 \\ 
-i\sigma_2\xi^{\dagger}   \end{array} \right ) \, .
\label{eq:weylc}
\end{eqnarray}
From this equation we can also read off
the relation between the 2-component Weyl fields and
their charge conjugates. Eq. (\ref{eq:weylc})
tells us  that $\psi^c_L(\psi^c_R)$ is a right-handed
(left-handed) Weyl field. Thus the  Weyl field operator $\psi_L(\psi_R)$
annihilates a Dirac fermion state $|\psi>$ having L (R) chirality and creates 
an antifermion state  $|{\bar\psi}>$ with R (L) chirality, while 
 $\psi^c_L(\psi^c_R)$  annihilates $|{\bar\psi}>$ having R (L)
chirality and creates a state $|\psi>$ with L (R) chirality. Moreover, we
immediately obtain that
\begin{eqnarray}
{\overline{\psi^c_L}} \: & \equiv & \: (\psi^c_L)^{\dagger}\gamma^0  \: = \: 
 (0, \:  i\eta^T\sigma_2 ) \, , \\ 
{\overline{\psi^c_R}}  \: & \equiv & \: (\psi^c_R)^{\dagger}\gamma^0  \: =\:
(-i\xi^T\sigma_2,\:  0) \, . 
\label{eq:weylcc}
\end{eqnarray}
\par
As to neutrinos,  there
are two options concerning their
nature (which must eventually be  resolved experimentally):
either Dirac or Majorana fermion.  The latter means, loosely
speaking,  that a neutrino
would be its own antiparticle. Actually, for a Majorana fermion
the distinction  between particle and antiparticle
looses its meaning because there is no longer  a conserved quantum 
number that would discriminate between them (see below). A Majorana field
is defined by the condition
\begin{equation}
\psi^c \: \stackrel{!}{=} \: r \psi \: ,
\label{eq:major}
\end{equation}
where $|r|=1$ is a phase chosen by convention. 
For  $r=+1$  the four-component  field
$ \psi_1 = (i\sigma_2\eta^{\dagger}, \, \eta )^T$ is a solution of 
this equation. In terms of Weyl fields this solution reads
\begin{equation}
\psi_1 \: =\: \psi_L \: + \: \psi^c_L \, .
\label{eq:pswli}
\end{equation}
The other solution of eq. (\ref{eq:major}) with $r=1$ is
\begin{equation}
\psi_2 \: =\: \psi_R \: + \: \psi^c_R \, .
\label{eq:pswre}
\end{equation}
Next we consider the Majorana mass terms. For  Majorana particles
described by $\psi_1$ and $\psi_2$ 
with masses  $m_1$ and $m_2$, respectively, 
we can write down  the following Majorana 
mass terms  
\begin{eqnarray}
{\cal L}^{(1)}_M \: & = & \: -\frac{m_1}{2}{\bar\psi_1}\psi_1 \: =\: 
-\frac{m_1}{2}{\overline{\psi^c_L}}\psi_L \: + \: {\rm h.c.} \: , \\
{\cal L}^{(2)}_M \: & = & \: -\frac{m_2}{2}{\bar\psi_2}\psi_2 \: =\: 
-\frac{m_2}{2}{\overline{\psi^c_R}}\psi_R  \: + \: {\rm h.c.} \:  ,
\label{eq:lmass12}
\end{eqnarray}
where we have used  that ${\bar\psi_A}\psi_A={\overline{\psi^c_A}}\psi^c_A=0$
for A=L,R. 
These mass terms violate the ``$\psi$-number''  by 2 units, 
$|\Delta{\rm L}_{\psi}|=2$. For instance 
$<{\bar\psi_R}|{\overline{\psi^c_L}}\psi_L|\psi_L>\neq 0$; i.e.,
the first term in 
${\cal L}^{(1)}_M$ flips a left-handed $|\psi_L>$ into a right-handed
$|{\bar\psi_R}>$. Recalling the connection 
between symmetries and conservation laws
we see that this non-conservation of $\psi$-number is related to the
fact that ${\cal L}^{(1,2)}_M$ are not invariant under the global
$U(1)$ transformation $\psi_{L,R}\to
e^{i\omega}\psi_{L,R},$  ${\bar \psi_{L,R}}\to
e^{-i\omega}{\bar\psi_{L,R}}$.
\par
The general mass term for  neutrino fields
$\nu_L$ and  $\nu_R$ contains both Majorana and  Dirac terms with
complex mass parameters. The 1-flavor
case reads 
\begin{eqnarray}
-{\cal L}_{D+M} \: & = & \: \frac{m_L}{2}{\overline{\nu^c_L}}\nu_L +
\frac{m_R}{2}{\overline{\nu^c_R}}\nu_R + m_D {\bar\nu_R}\nu_L \: + \: 
{\rm h.c.}\\
& = &  \frac{1}{2}({\bar\psi_1},\,{\bar\psi_2})
\left (\begin{array}{cc} m_L & m_D \\ 
m_D & m_R  \end{array} \right ) \, \left (\begin{array}{c} \psi_1 \\
\psi_2 \end{array} \right ) \: \, ,
\label{eq:dipma}
\end{eqnarray}
where 
\begin{eqnarray}
\psi_1 \: & = & \: \nu_L  \: + \: {\nu^c_L} \, , \\
\psi_2 \: & = & \: \nu_R  \: + \: {\nu^c_R} 
\label{eq:maeif}
\end{eqnarray}
are Majorana fields. The  mass parameters in (\ref{eq:dipma})
 are taken to be real.
 Let's diagonalize the mass matrix for  the
case $m_R>>m_D>>m_L$.  Putting  $m_L=0$ we have in the mass basis
\begin{equation}
-{\cal L}_{D+M} \:  = \: \frac{m_{\nu}}{2}{\bar\nu}\nu
\:+ \: \frac{m_{N}}{2}{\bar N}N \, ,
\label{eq:mama}
\end{equation}
where
\begin{equation}
\nu \:  \simeq \: \psi_1 \; , \hspace*{0.5cm}
N \:  \simeq \: \psi_2 \, , 
\label{eq:mafie}
\end{equation}
and 
\begin{eqnarray}
- m_{\nu} \: & \simeq & \: \frac{m_D^2}{m_R} \, << \, m_D \,  , 
\label{eq:mawal1} \\
m_N \: & \simeq & \: m_R \,. 
\label{eq:mawal2}
\end{eqnarray}
The eigenvalue $m_{\nu}$ can be made positive by an appropriate
change of phase of the field $\nu$. 
For $m_R>>m_D$ the neutrino mass eigenstates consist of a 
very light left-handed state $|\nu>$ and a very heavy right-handed
state  $|N>$. Eq. (\ref{eq:mawal1}) constitutes the 
seesaw mechanism \cite{seesaw} for generating a very small mass for a
left-handed neutrino 
from $m_D={\cal O}(h_{\ell}v)$ and a large $m_R$.
\par
Finally we consider the case of 3 lepton generations. Denoting
the $SU(2)_L$ doublets $\ell \equiv (\nu_{\alpha L}, \ell_{\alpha L})^T$,  
and the $SU(2)_L$ singlets $e_R \equiv \ell_{\alpha R}$,  the
 $SU(2)_L \times U(1)_Y$ singlets 
$\nu_R \equiv \nu_{\alpha R},$ 
where $ \alpha=e,\mu,\tau$ labels the lepton generations in the weak basis and
$\tilde\Phi_r \equiv i\sigma_2\Phi_r^\dagger$, $r=1,2$, 
where $\Phi_r$ are  Higgs doublet fields, the general
 $SU(2)_L \times U(1)_Y$ invariant Yukawa interactions in the lepton
sector read 
\begin{equation}
-{\cal L}_{Y} =  {\bar \ell_L} \Phi_1 h_e e_R + {\bar \ell_L} 
{\tilde\Phi_2} h_{\nu} \nu_R +\frac{1}{2} {\overline{\nu^c_R}} M_R \nu_R \: + \:
h.c. \: .
\label{eq:3gen}
\end{equation}
Here $h_e, h_{\nu}$ denote the complex, $3\times 3$ Yukawa coupling
 matrices, and $M_R$ is the  $3\times 3$ mass matrix for the right-handed
neutrino fields which may be taken to be diagonal
without loss of generality. ($M_R$ can be generated
by a large VEV of a gauge singlet Higgs field.) Spontaneous symmetry
breaking at the electroweak phase transition, $<0|\Phi_r|0>_T =v_{rT}/\sqrt{2}$,
leads to  Dirac mass matrices for the charged leptons and
neutrinos,
\begin{equation}
M_e = h_e \frac{v_{1T}}{\sqrt{2}}, \hspace*{0.5cm}
 M_{D} = h_{\nu} \frac{v_{2T}}{\sqrt{2}} \; .
\label{eq:D3gen}
\end{equation}
Let us  change from the weak basis to the mass basis by performing
appropriate
unitary transformations in flavor space. 
Using that the matrix elements of $M_R$ are much larger than those
of $M_D$ one obtains \cite{Buchmuller:2000as}
\begin{eqnarray}
\nu_i \: & \simeq & \: (K^\dagger)_{i\alpha}\nu_{\alpha L}
\: +  \:  \nu_{\alpha L}^c K_{\alpha i} \,  , \\
N_i \: & \simeq & \:   \nu_{\alpha R} \: +  \:  \nu_{\alpha R}^c \: ,  
\label{eq:ms3g}
\end{eqnarray}
with the diagonal mass matrices
\begin{eqnarray}
M_{\nu} \:  & = & - \:  K^{\dagger} M_D M_R^{-1}M_D^TK^*
\, +\, {\cal O}(M_R^{-3})\,  ,
\label{eq:ms3m1} \\
M_N\: &  =  & \: M_R \, +\, {\cal O}(M_R^{-1})\: , 
\label{eq:ms3m}
\end{eqnarray}
where $i=1,2,3$ labels the fields in the mass basis and $K$ is the
unitary $3\times 3$ matrix which describes the mixing of
 the lepton flavors
in the charged current interactions 
\begin{equation}
{\cal L}_{cc}^{lept} = - \frac{g_w}{\sqrt{2}} 
 {\bar \ell_{iL}}\gamma^{\mu}K_{ij}\nu_{j}W^-_{\mu} \: + \:
h.c. \; .
\label{eq:lcclep}
\end{equation}
We can decompose the Dirac mass matrix $M_D$ into the form
\begin{equation}
M_D   \:  =   \: V R U^{\dagger}  \; ,
\label{eq:mddec}
\end{equation}
where $U,V$ are unitary matrices and $R=diag(r_1,r_2,r_3)$. From (\ref{eq:ms3m1}) it follows that
the moduli and phases of the matrix elements of
$K$, which are relevant for present-day neutrino physics --
e.g., for neutrino oscillations or for the search for neutrinoless
double beta decay $Z\to (Z+2)+2e^-$ -- depend on the mass ratios
$m_j/m_i$ of the light, left-handed neutrinos, and on the angles and
phases of $U$ {\it and}   $V$. On the other hand the matrix
$M_D^\dagger M_D$, on which the quantities responsible
for leptogenesis, in particular  the CP asymmetry depend (see section 6.2),
is given by 
\begin{equation}
M_D^\dagger M_D   \:  =   \: U R^2 U^\dagger  \; .
\label{eq:md2dec}
\end{equation}
Hence for leptogenesis only the CP-violating phases of $U$ are relevant!
Therefore, in this scenario there is in general no connection between
possible CP-violating effects that could be traced in the laboratory
and the CP-violating phases which  are responsible for the generation of the
BAU.

\end{document}